\documentclass[amsmath,twocolumn,prb,nofootinbib,longbibliography,superscriptaddress]{revtex4-2}

\usepackage{graphicx}
\usepackage{dcolumn}
\usepackage{bm}
\usepackage{braket}
\usepackage{xcolor} 
\usepackage{hyperref}
\usepackage{soul}
\usepackage{amssymb}

\begin{document}

\title{Charge density wave induced gapped nodal line}

\author{Sergey Alekseev}
\affiliation{Department of Physics and Astronomy, Stony Brook University, Stony Brook, NY 11794}
\author{Lei Chen}
\affiliation{Department of Physics and Astronomy, Stony Brook University, Stony Brook, NY 11794}
\author{Jennifer Cano}
\affiliation{Department of Physics and Astronomy, Stony Brook University, Stony Brook, NY 11794}
\affiliation{Center for Computational Quantum Physics, Flatiron Institute, New York, NY 10010}

\begin{abstract}
We investigate the interplay between charge density wave (CDW) order and topological nodal-line states in square-net materials. 
Our Ginzburg-Landau theory predicts a CDW instability that generically opens a gap at the Fermi energy while preserving the nodal line crossing.
However, as the Fermi level approaches the nodal line, the density of states at the nodal line decreases, eventually disappearing as the CDW vector $\mathbf{Q}$ goes to zero.
Exactly at $\mathbf{Q} = 0$, the order parameter explicitly breaks the glide symmetry protecting the nodal line, which allows a gap to open.
Yet, for small but finite $\mathbf{Q}$, 
the nodal line may vanish within experimental resolution even when the glide symmetry is preserved.
Our results provide a consistent explanation for recent experimental observations.
\end{abstract}

\maketitle

\section{\label{sec:introduction} Introduction}

Charge density waves (CDWs) have been extensively studied in the square-net rare-earth tritellurides ($R$Te$_3$, $R$ = rare-earth element) \cite{dimasi1995chemical,brouet2004Fermi,komoda2004high,laverock2005Fermi,kim2006local,malliakas2006divergence,fang2007stm,brouet2008angle,ru2008effect, Eiter_2012,sinchenko2014unidirectional,maschek2015wave,kogar2019light,zong2019evidence,walmsley2020magnetic,sharma2020interplay,liu2020electronic,zhou2021nonequilibrium,gonzalez2022time,wang2022axial,straquadine2022evidence,Singh2023,Kivelson2023,chikina2023charge,raghavan2023atomic,Kim2024,singh2024ferro,akatsuka2024noncoplanar}. In these materials, the CDW transition is driven by electrons originating primarily from the two-dimensional (2D) Te square nets. The partially filled $p$-orbitals in these layers form quasi-one-dimensional Fermi surfaces that exhibit nearly perfect nesting \cite{dimasi1995chemical,johannes2008fermi}, which enhances the electronic susceptibility at the nesting wave vector and promotes a CDW instability \cite{kivelson2006stripes}. 
Though CDWs in this family have been studied for several decades, recent experiments continue to reveal new aspects, including ferroaxial order arising from unconventional CDW order~\cite{wang2022axial,singh2024ferro,2024orbitaltextures}.

The square net motif also gives rise to a celebrated nodal line protected by the nonsymmorphic crystal symmetry \cite{schoop2016dirac,klemenz2020systematic}.
The nodal line is characterized by a topological invariant associated with drumhead surface states \cite{burkov2011topological,chiu2014classification,muechler2020modular}.
However, so far, the nodal line has not been studied in the $R$Te$_3$ family, where it may be far from the Fermi level or buried in other bands. 

In contrast, the nodal line has been a topic of intense recent interest in the related $R$SbTe family \cite{xu2015twodimensional,hosen2018discovery,schoop2018tunable,yang2020magnetic,yue2020topological,wang2021spectroscopiceveidencenodallinesemimetal,pandey2021magnetic,regmi2022observation,yuan2024observation}; in particular, LaSbTe has been reported as a genuine nodal line semimetal where the nodal crossing nearly coincides with the Fermi level \cite{xu2015twodimensional,wang2021spectroscopiceveidencenodallinesemimetal}.
This family exhibits a similar crystal structure to the $R$Te$_3$ compounds, with Sb square nets between $R$Te layers.
The similar structure results in similar Fermi surface nesting, which may also gives rise to a CDW instability \cite{dimasi1996stability,schoop2019cdwandmagnetism,lei2021band,lei2021complex,singha2021evolving,li2021charge,schoop2023ultrafast,que2025visualizing}.
Thus, the $R$SbTe family provides a unique opportunity to study the interplay between CDWs and topological nodal line semimetals; indeed, earlier work proposed CDWs as a route to engineer the band structure of topological semimetals \cite{lei2021band}.

A recent experiment on LaSb$_x$Te$_{2-x}$~\cite{bannies2024electronicallydrivenswitchingtopologylasbte} demonstrated that the nodal line could be controllably gapped and restored \textit{in situ} by electron doping. 
The gap is attributed to a structural phase transition and accompanying CDW that breaks the glide symmetry protecting the nodal crossing.
However, the existence of a CDW by itself is insufficient to gap the nodal loop, for two reasons: first, generically the chemical potential does not sit exactly at the nodal line and hence the gap opened by the CDW may be well-separated 
from the energy of the nodal line; 
and second, the CDW may preserve an exact or approximate glide symmetry of the enlarged CDW-modulated supercell, which can also protect the nodal line crossing. Indeed, the calculations and measurements in Ref.~\cite{lei2021band} show that the nodal line in GdSb$_x$Te$_{2-x}$ is preserved in the CDW phase.
Thus, Ref.~\cite{bannies2024electronicallydrivenswitchingtopologylasbte} raises an important question as to the mechanism by which a CDW can gap a symmetry-protected nodal line; answering this question is the purpose of the present study.

In this work, we develop a theoretical model to understand the conditions under which a CDW can open a gap along the nodal line in square-net materials.
Specifically, we derive a Ginzburg–Landau (GL) theory to describe the CDW transition and analyze the theory as a function of chemical potential to model the electron doping. 
We find that more important than whether the CDW breaks or preserves the glide symmetry is how close the chemical potential lies to the nodal line:
when the chemical potential is far from the nodal line, the onset of the CDW opens gaps above and below the nodal line while preserving the nodal crossing, consistent with Ref.~\cite{lei2021band}.
As the chemical potential approaches the nodal line, the nesting vector $\mathbf{Q}$ shrinks, and the density of states near the nodal line decreases monotonically, eventually becoming fully gapped when $\mathbf{Q} = 0$.
The nature of the CDW is such that the $\mathbf{Q} = 0$ limit corresponds to a distortion that breaks the symmetry between the square-net sublattices, thereby breaking the glide symmetry.
For sufficiently small but nonzero $\mathbf{Q}$, even though a glide symmetry of the supercell may be present, the density of states associated with the nodal line can be drastically suppressed and becomes nearly invisible in, e.g., angle-resolved photoemission spectroscopy (ARPES) measurements. 
The trend that we observe in the density of states at the nodal line as a function of electron doping is consistent with the evolution of the nodal-line gap observed in Ref. \cite{bannies2024electronicallydrivenswitchingtopologylasbte}. 

The structure of the paper is as follows. Sec.~\ref{sec:model} introduces the tight-binding model of a square net with two sublattices and Sec.~\ref{sec:symmetries} discusses the symmetries of the model. In Sec.~\ref{sec:order_parameters}, we define the CDW order parameters and investigate their transformation properties under crystallographic symmetries.
This analysis forms the basis for deriving and solving the GL theory in Sec.~\ref{sec:GL_theory}. Sec.~\ref{sec:mft_theory} then examines how the CDW order parameter can open a gap in the nodal line. In Sec.~\ref{sec:experiment}, we discuss the experimental implications of our findings. Finally, we conclude our work in Sec.~\ref{sec:conclusion}.

\section{\label{sec:model} Microscopic Model}
\begin{figure}
    \centering
    \includegraphics[height=4cm]{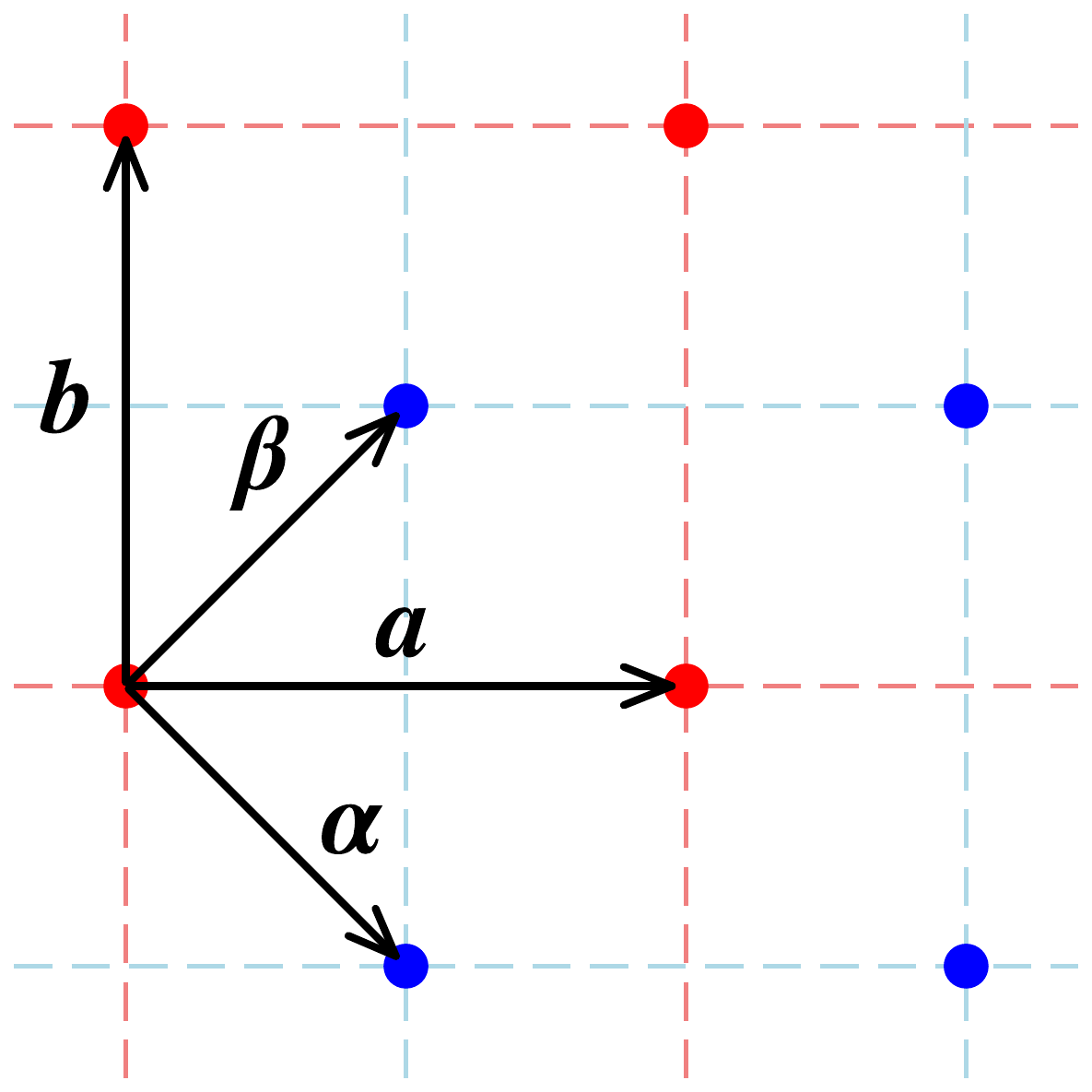}
    \includegraphics[height=4.2cm]{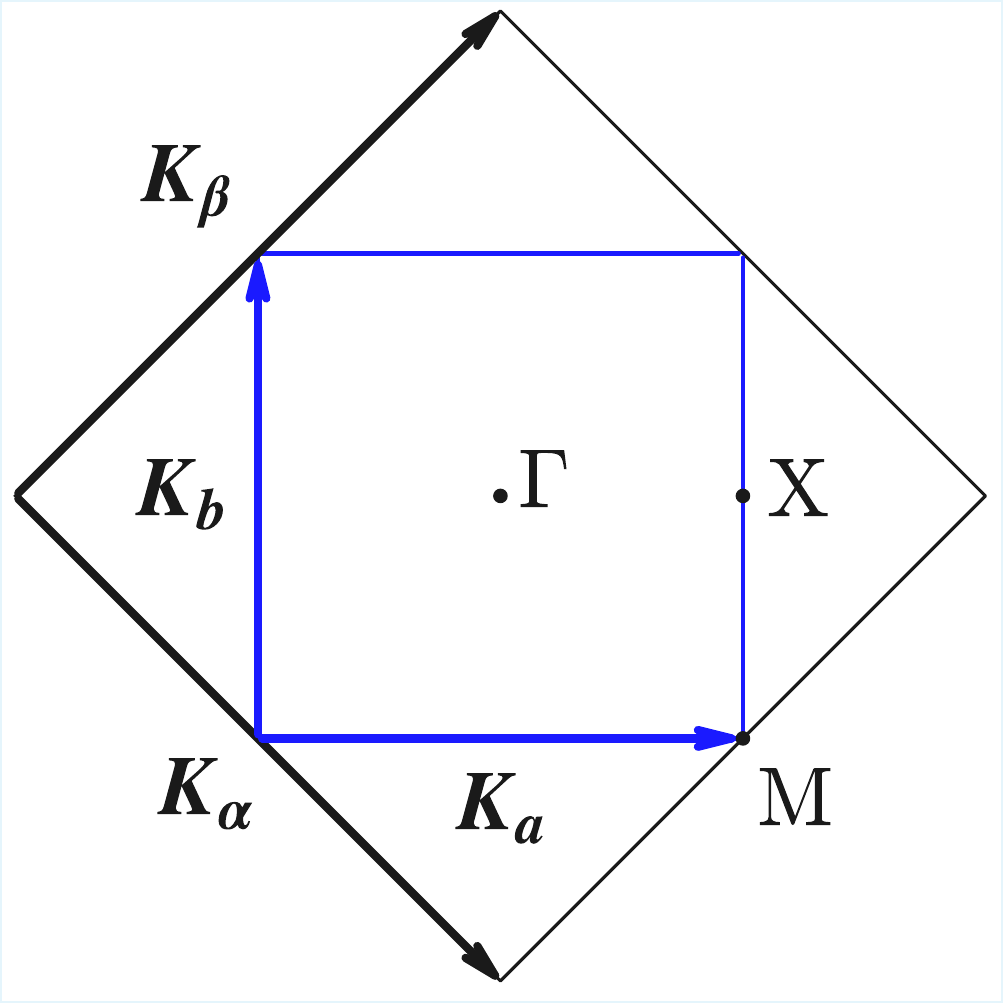}
{\includegraphics[height=3.8cm]{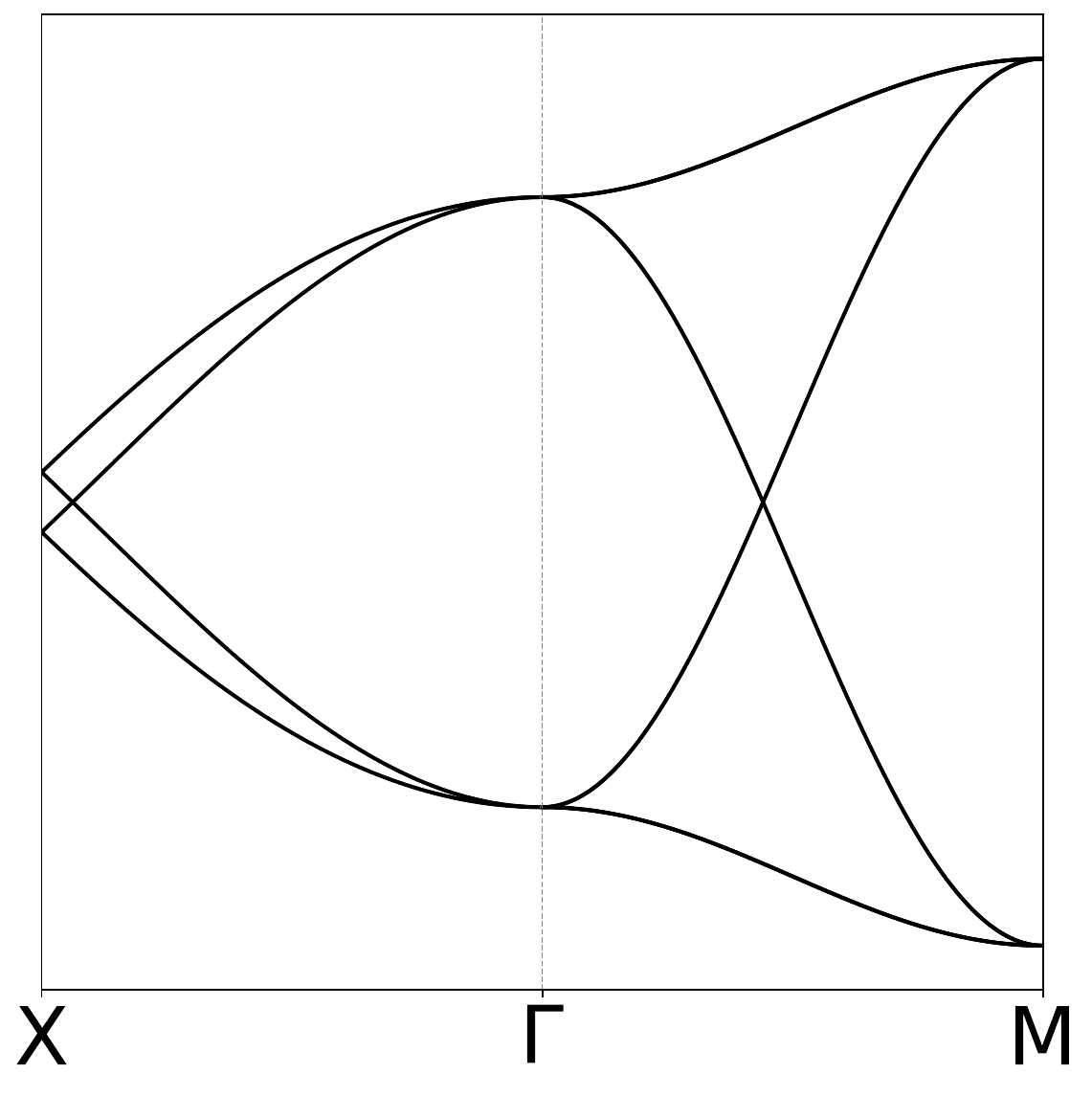}}
    \caption{
    Upper left panel: Square lattice with two sublattices. The lattice vectors $\bm{\alpha}$ and $\bm{\beta}$ define the unit cell of the original 2D square net. The vectors $\bm{a}= (1,0)$ and $\bm{b}= (0,1)$ define the enlarged unit cell, obtained by projecting the unit cell of the bulk crystal onto the 2D square net.
    Upper right panel: BZ before (black) and after (blue) folding, with corresponding reciprocal lattice vectors indicated. 
    Lower panel: Energy bands of the Hamiltonian in Eq. (\ref{eq: 4_4_ham}) along the X-$
    \Gamma$-M  path. The band crossings are part of a nodal line in the 2D plane.
    }
    \label{fig:lattice}
\end{figure}

The low-energy electronic structure of square-net materials such as in the $R$SbTe or $R$Te$_3$ families arises primarily from the $p_x$ and $p_y$ orbitals of Te or Sb atoms, which form a two-dimensional (2D) square lattice.
Due to the atomic structure of the layers above and below this square net, the unit cell of the bulk crystal projected onto the 2D layer is twice as big as the unit cell of the single layer, i.e., it forms a $\sqrt{2} \times \sqrt{2}$ square lattice relative to the original. The original lattice vectors $\bm{\alpha}$ and $\bm{\beta}$, along with the lattice vectors $\bm{a}=(a,0)$ and $\bm{b}=(0,a)$ of the enlarged unit cell, are shown in Fig.~\ref{fig:lattice}; we use units where $a=1$. In the following, we derive a Hamiltonian for the square net by first assuming the atoms in the square net are equivalent and then folding the band structure into the Brillouin zone (BZ) of the $\sqrt{2} \times \sqrt{2}$ unit cell.

To this end, we start with a square lattice with $p_x$ and $p_y$ orbitals on each site.
Including only nearest- and next-nearest-neighbor hopping, the Hamiltonian takes the following form, written in the orbital basis:
\begin{widetext}
\begin{equation}
\tilde h_{\bm{k}}=
\begin{pmatrix}
2t_\sigma \cos (\bm{k}\cdot \bm{\alpha}) -2t_\pi \cos (\bm{k}\cdot \bm{\beta})  & 2 t_d \sin (\bm{k}\cdot \bm{\alpha}) \sin (\bm{k}\cdot \bm{\beta}) \\
2 t_d \sin (\bm{k}\cdot \bm{\alpha}) \sin (\bm{k}\cdot \bm{\beta}) & 2t_\sigma \cos (\bm{k}\cdot \bm{\beta})-2t_\pi \cos (\bm{k}\cdot \bm{\alpha})
\end{pmatrix} \equiv 
\begin{pmatrix}
a_{\bm{k}} & c_{\bm{k}} \\
c_{\bm{k}} & b_{\bm{k}}
\end{pmatrix},
\label{eq: 2_2_ham}
\end{equation}
\end{widetext}
where $\bm{k}$ belongs to the BZ defined by the reciprocal lattice vectors $\bm{K}_{\bm{\alpha}}$ and $\bm{K}_{\bm{\beta}}$. In what follows, the hopping amplitudes are taken as $t_\sigma \approx 2.0$~eV, $t_\pi \approx 0.37$~eV, and $t_d \approx 0.16$~eV, as estimated for a rare-earth tellurium square net~\cite{kivelson2006stripes}.
A square net composed of a different element may change these parameters, but we do not expect it to qualitatively change our results.
Introducing the Pauli matrices $\sigma_i$ $(i=1,2,3)$
and the identity matrix $\sigma_0$ that act in the orbital space, we rewrite the Hamiltonian as:
\begin{equation}
\tilde h_{\bm{k}}=
\frac{a_{\bm{k}} + b_{\bm{k}}}{2} \sigma_0 + \frac{a_{\bm{k}} - b_{\bm{k}}}{2} \sigma_z + c_{\bm{k}}\, 
\sigma_x.
\label{eq:2_2_ham_pauli}
\end{equation}

We now fold the Hamiltonian into the reduced BZ spanned by the reciprocal lattice vectors $\bm{K_a}$ and $\bm{K_b}$, which correspond to real-space lattice vectors $\bm{a}$ and $\bm{b}$, as shown in Fig.~\ref{fig:lattice}.
The 
enlarged unit cell has two atoms, which define two sublattices that we call $A$ and $B$ (as indicated by red and blue sites in Fig.~\ref{fig:lattice}).
To derive the folded Hamiltonian, note that the first two terms in Eq.~(\ref{eq:2_2_ham_pauli}) describe hopping between atoms on different sublattices, while the last term corresponds to hopping between atoms within the same sublattice.
We thus arrive at the band-folded Hamiltonian:
\begin{equation}
    h_{\bm{k}} = \frac{a_{\bm{k}} + b_{\bm{k}}}{2} \sigma_0 \otimes \tau_x + \frac{a_{\bm{k}} - b_{\bm{k}}}{2} \sigma_z \otimes \tau_x + c_{\bm{k}}\, \sigma_x \otimes 
    \tau_0.
\label{eq: 4_4_ham}
\end{equation}
Here, we have introduced another set of Pauli matrices $\tau_i$ $(i=1,2,3)$ and $\tau_0$, the identity matrix, which act on the sublattice indices. 
Now, specifying an atom in the lattice requires specifying not only its unit cell, given by the vector $\bm{R} = n\bm{a} + m\bm{b}$, but also its sublattice type ($A$ or $B$).
Due to the enlarged unit cell, $\bm{k}$ in Eq.~(\ref{eq: 4_4_ham}) now belongs to the folded BZ defined by the reciprocal lattice vectors $\bm{K}_{\bm{a}}$ and $\bm{K}_{\bm{b}}$.

Below, we will study a CDW order parameter that arises from spontaneous symmetry breaking induced by an on-site attractive ($g>0$) interaction: 
\begin{equation}
V = - \frac{g}{2}
    \sum_{\substack{{\bm{k}, \bm{k'}, \bm{Q}} \\ {\alpha\beta,I} }} \psi^\dagger_{\bm{k}+\bm{Q},\alpha I} \psi_{\bm{k},\beta I}\ \psi^\dagger_{\bm{k'}-\bm{Q}, \beta I}\psi_{\bm{k'},\alpha I},
    \label{eq:coulomb_interaction}
\end{equation}
where $\alpha, \beta$ are the orbital indices $p_x, p_y$; $I \in \{A,B\}$ is the sublattice index; and the sum over $\bm{k}, \bm{k'},\bm{Q}$ runs over the folded Brillouin zone.
Since the $A$ and $B$ sublattices are separated in real space, the on-site interaction does not couple the sublattices.

\section{\label{sec:symmetries} Symmetries}

We now discuss the symmetries of our model. 
The symmetry of a two-dimensional square lattice is described by the wallpaper group $p4mm$, with the corresponding point group $D_{4h}$. As mentioned in the introduction, the bulk crystal exhibits a glide symmetry. 
In the square net layer this symmetry is implemented by a fractional translation with respect to the enlarged $\sqrt{2}\times\sqrt{2}$ unit cell that exchanges the sublattices. For example, a translation by $\bm{\alpha}$ acts on the fermion operators in unit cell  $\bm{R}$ by: $\psi_{\bm{R},A} \mapsto \psi_{\bm{R},B}$, $\psi_{\bm{R},B} \mapsto \psi_{\bm{R}+2\bm{\alpha},A}$. In momentum space, this transformation becomes
\begin{equation}
\bm{\alpha}:\quad
\psi_{\bm{k},A(B)} \mapsto e^{i\bm{k}\cdot \bm{\alpha}} \psi_{\bm{k},B(A)},
\label{eq:glide}
\end{equation}
using the Fourier convention $\psi_{\bm{R},A} = \sum_{\bm{k}} e^{i\bm{k}\cdot\bm{R}} \psi_{\bm{k},A}$, $\psi_{\bm{R},B} = \sum_{\bm{k}} e^{i\bm{k}\cdot(\bm{R} +\bm{\alpha})} \psi_{\bm{k},B}$. 
Since this symmetry descends from the bulk glide symmetry, we refer to it as a glide symmetry, even though it appears as a (fractional) translation symmetry within our square net model.
Note that we could have equivalently chosen to implement the glide symmetry by translation by $\bm{\beta}$.

The glide symmetry protects the band crossings in the band structure of the folded Hamiltonian (\ref{eq: 4_4_ham}), as illustrated in Fig. \ref{fig:lattice}. These crossings form a nodal loop that extends across the entire Fermi surface when the bands are half-filled.
Though our tight-binding model only includes nearest- and next-nearest hopping, the nodal line is a robust feature that would remain gapless 
even if longer-range hopping was included in the Hamiltonian precisely because it is protected by the glide symmetry.

In Sec.~\ref{sec:mft_theory}, we discuss the conditions under which a glide symmetry of the CDW supercell is exactly or approximately preserved even when the glide symmetry defined in Eq.~(\ref{eq:glide}) is broken. Such a supercell glide symmetry also protects the nodal line.

\section{\label{sec:order_parameters}CDW Order Parameters}
We introduce separate CDW order parameters for each sublattice, $\Delta^A_{\bm{Q}}$ and $\Delta^B_{\bm{Q}}$ :
\begin{equation}
    \Delta^{I}_{\bm{Q}} = \sum_{\bm{k}}
    \Braket{
    \psi_{\bm{k-Q}, I}^\dagger
    \psi_{\bm{k}, I}
    },\ I \in \{A,B\},
    \label{eq: ab_order_params_def}
\end{equation}
where $\bm{k}$ runs over the folded BZ.
For simplicity, we assume that both order parameters are trivial in orbital space and omit summation over repeated orbital indices. Nontrivial orbital order could give rise to a coexisting ferroaxial phase \cite{2024orbitaltextures}, but such a study is beyond the scope of the present work.

The order parameters (\ref{eq: ab_order_params_def}) have the following transformation properties under symmetries $g$ belonging to the point group $D_{4h}$, translations $\bm{t}$, and the glide transformation $\bm{\alpha}$:
\begin{eqnarray}
    && g:\quad \Delta^{A(B)}_{\bm{Q}} \mapsto \Delta^{A(B)}_{g^{-1}\cdot \bm{Q}}, \nonumber \\
    && \bm{t}: \quad
    \Delta^{A(B)}_{\bm{Q}} \mapsto e^{i \bm{t}\cdot \bm{Q}}  \Delta^{A(B)}_{\bm{Q}}, \nonumber \\
    && \bm{\alpha}: \quad
    \Delta^{A(B)}_{\bm{Q}} \mapsto e^{i \bm{\alpha} \cdot \bm{Q}} \Delta^{B(A)}_{\bm{Q}}.
\end{eqnarray}
Therefore, the following two linear combinations form irreps under glide symmetry:
\begin{eqnarray}
    && \Delta^0_{\bm{Q}} = \Delta^A_{\bm{Q}} + \Delta^B_{\bm{Q}} \mapsto + e^{i \bm{\alpha} \cdot \bm{Q}} \Delta^0_{\bm{Q}}, \nonumber\\
    && \Delta^z_{\bm{Q}} = \Delta^A_{\bm{Q}} - \Delta^B_{\bm{Q}} \mapsto  -e^{i \bm{\alpha} \cdot \bm{Q}} \Delta^z_{\bm{Q}}.
\label{eq:glide_transformation_phases}
\end{eqnarray}

As noted below Eq.~(\ref{eq: ab_order_params_def}), $\mathbf{Q}$ is restricted to the folded BZ, which prevents a redundancy in the description of the order parameters. Specifically, 
since $\bm{\alpha} = (\bm{a} - \bm{b})/2$, which implies  $\bm{K}_{\bm{a}} \cdot \bm{\alpha} = \pi$, 
\begin{eqnarray}
    && \Delta^{0}_{\bm{Q} + \bm{K}_{\bm{a}}} \mapsto -e^{i \bm{\alpha} \cdot \bm{Q}} \Delta^0_{\bm{Q} + \bm{K}_{\bm{a}}}, \nonumber \\
    && \Delta^{z}_{\bm{Q} + \bm{K}_{\bm{a}}} \mapsto + e^{i \bm{\alpha} \cdot \bm{Q}} \Delta^z_{\bm{Q} + \bm{K}_{\bm{a}}}.
\end{eqnarray}
Thus, the two order parameters $\Delta^0_{\bm{Q}}$ and $\Delta^z_{\bm{Q} + \bm{K}_{\bm{a}}}$ describe the same physical phase. 
This redundancy follows from the equivalence between $\bm{Q}$ and $\bm{Q}+\bm{K}_{\bm{a}}$ after band folding. 

The equivalence of $\Delta^0_{\bm{Q}}$ and $\Delta^z_{\bm{Q} + \bm{K}_{\bm{a}}}$ is also apparent in real space. In the ``$\Delta^0$-phase'', where $\Delta^0_{\bm{Q}} \neq 0$ and $\Delta^z_{\bm{Q}} =0$, which implies $\Delta^A_{\bm{Q}} = \Delta^B_{\bm{Q}}$, the real space order parameters are given by:
\begin{eqnarray}
    && \Delta^A(\bm{R}) = \cos \left( \bm{Q} \cdot \bm{R} \right), \nonumber \\
    && \Delta^B(\bm{R}) = \cos \left( \bm{Q} \cdot \bm{R} + \bm{Q} \cdot \bm{\alpha} \right)
\label{eq: d0_phase_real_space}
\end{eqnarray}
using the Fourier convention below Eq. (\ref{eq:glide}).
For brevity, in the formula above and in subsequent expressions, we omit any common amplitude and phase. In the ``$\Delta^z$-phase'', where $\Delta^z_{\bm{Q}} \neq 0$ and $\Delta^0_{\bm{Q}} =0$, it follows that $\Delta^A_{\bm{Q}} = -\Delta^B_{\bm{Q}}$. In real space, this translates to:
\begin{eqnarray}
    && \Delta^A(\bm{R}) = \cos \left( \bm{Q} \cdot \bm{R} \right), \nonumber \\
    && \Delta^B(\bm{R}) = - \cos \left( \bm{Q} \cdot \bm{R} + \bm{Q} \cdot \bm{\alpha} \right).
\label{eq: dz_phase_real_space}
\end{eqnarray}
Comparing Eqs.~(\ref{eq: d0_phase_real_space}) and Eqs.~(\ref{eq: dz_phase_real_space}) shows that the $\Delta^z$-phase at wave vector $\bm{Q} = \bm{Q}'+\bm{K_a}$ corresponds to an identical charge density modulation as the $\Delta^0$-phase at $\bm{Q}'$.
This follows from the fact that $\bm{R}$ is a lattice vector of the enlarged unit cell, so $\bm{K}_{\bm{a}}\cdot \bm{R}$ is a multiple of $2\pi$,
while $\bm{K_a} \cdot \bm{\alpha} = \pi$.
\begin{figure}
    \centering
    \hspace{-1.4cm}
    \vspace{0.3cm}
    \includegraphics[height=6.6cm]{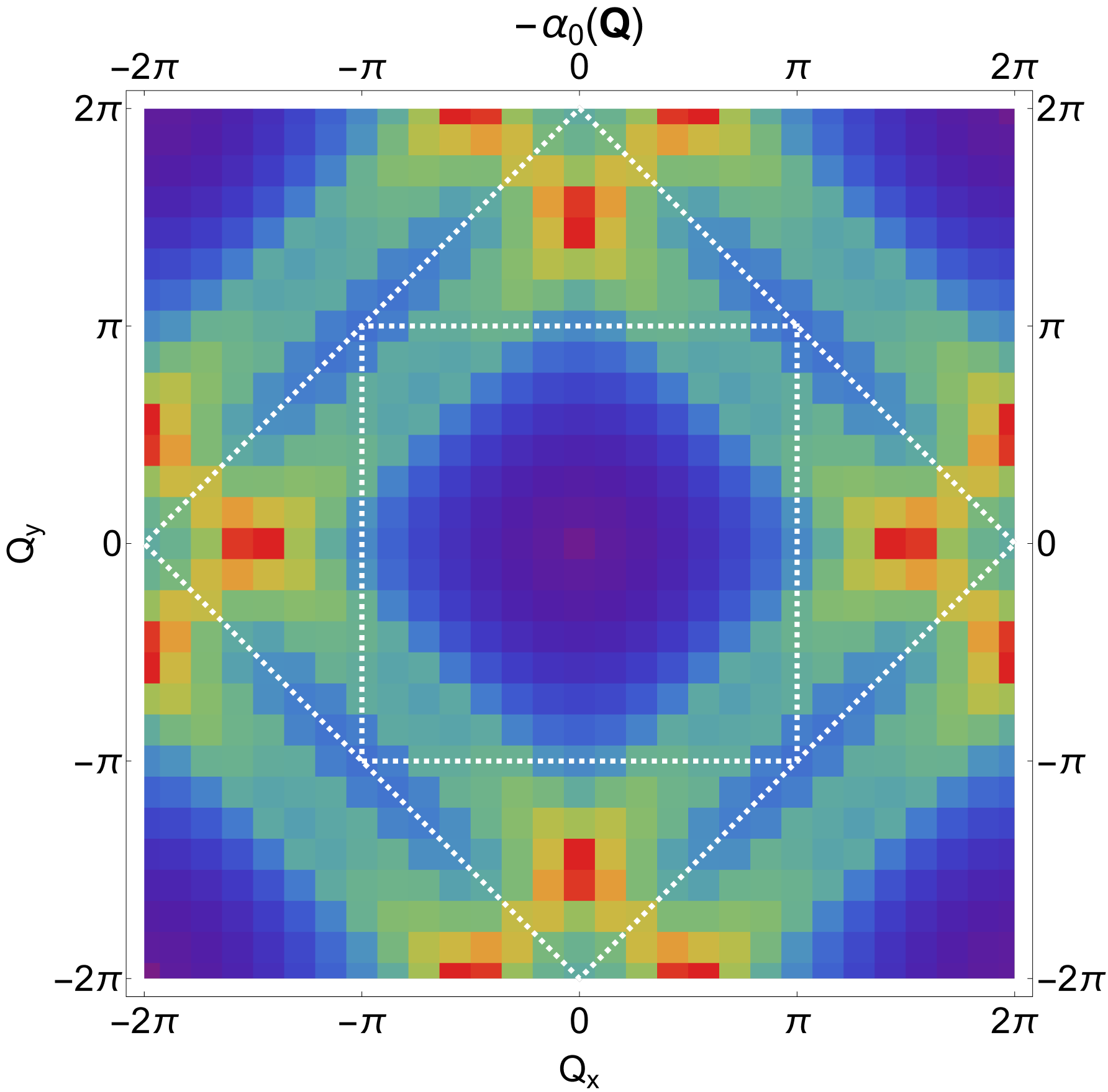}
    \includegraphics[height=6.6cm]{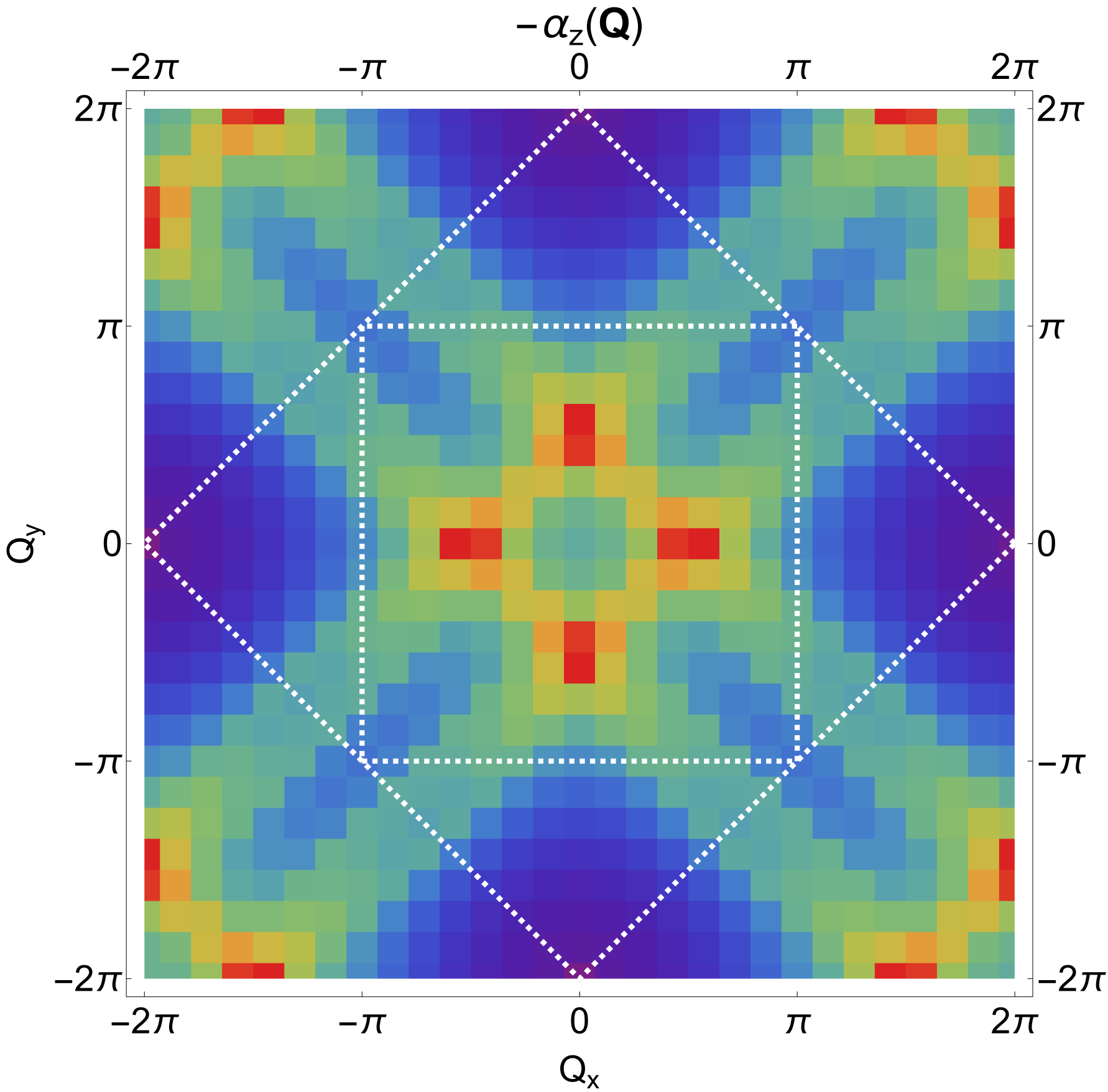}
    \hspace{0.2cm}
    \raisebox{3.5cm}[0pt][0pt]{\includegraphics[height=5.8cm]{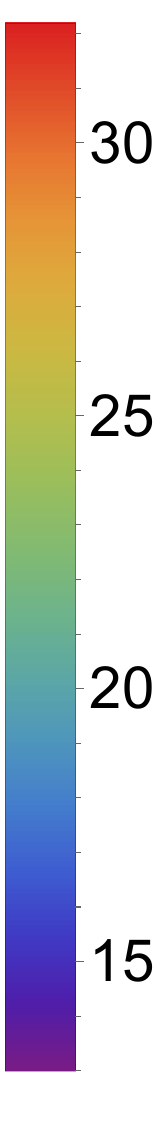}}
    \caption{The quadratic coefficients in the GL theory, $-a_0(\bm{Q})$ (upper panel) and $-a_z(\bm{Q})$ (lower panel), as functions of $\bm{Q}$, at fixed temperature $T=0.03$~eV and chemical potential $\mu=1.53$~eV, in units where the lattice constants $|\mathbf{a}| = |\mathbf{b}|=1$. The white dashed squares indicate the folded and unfolded Brillouin zones introduced in Fig.~\ref{fig:lattice}.
    }
    \label{fig:a0_az_vs_q_2d}
\end{figure}
\begin{figure*}
    \centering
    \includegraphics[height=10.0cm]{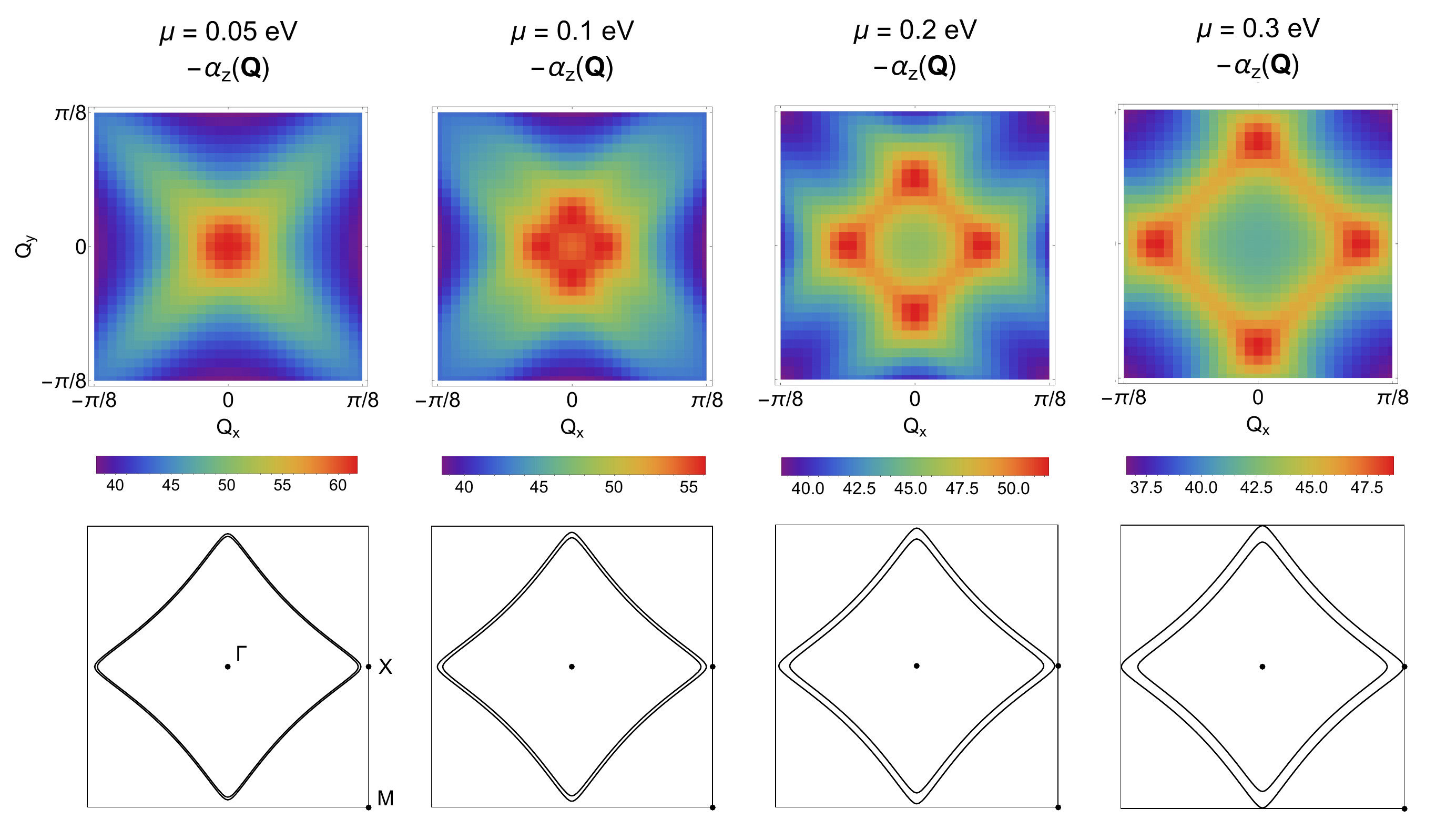}

    \caption{
The quadratic coefficient in the GL theory $-a_z(\bm{Q})$ as a function of $\bm{Q}$ at fixed temperature $T=0.03$~eV and multiple values of chemical potential, along with the corresponding Fermi surfaces.
    } 
    \label{fig:az_vs_q_smaller_mu}
\end{figure*}

\section{\label{sec:GL_theory} Ginzburg-Landau theory}
The Ginzburg-Landau (GL) theory is built from combinations of the order parameters $\Delta^{0,z}_{\bm{Q}}$ that form translation-invariant bilinears $\Delta^i_{\bm{Q}} \Delta^j_{-\bm{Q}}$ ($i,j=0,z$).
We showed in Ref.~\cite{2024orbitaltextures} that the
Lindhard susceptibility is peaked when the CDW vector $\bm{Q}$ aligns with the lattice vector $\bm{a}$.
Assuming a single-$\bm{Q}$ CDW, we thus restrict our analysis to the crystallographic point group $D_{2h}$, which consists of symmetries that either preserve $\bm{Q}$ or map $\bm{Q} \rightarrow -\bm{Q}$. 
The allowed bilinears decompose into the following irreps under $D_{2h}$ and glide symmetry:
\begin{eqnarray}
    && A_{g},\ +1:
    \quad
    |\Delta^0_{\bm{Q}}|^2,\ |\Delta^z_{\bm{Q}}|^2; \nonumber \\
    && A_{g},\ -1:
    \quad 
    \overline {\Delta}^{0}_{\bm{Q}}\Delta^z_{\bm{Q}} + \Delta^0_{\bm{Q}} \overline {\Delta}^{z}_{\bm{Q}};
    \nonumber \\
    && B_{3u},\ -1:
    \quad
    \overline {\Delta}^{0}_{\bm{Q}}\Delta^z_{\bm{Q}} - \Delta^0_{\bm{Q}} \overline {\Delta}^{z}_{\bm{Q}},
\label{eq:bilinears_irreps}
\end{eqnarray}
where we have used the relation $\Delta^{0,z}_{-\bm{Q}} = \overline {\Delta}^{0,z}_{\bm{Q}}$, with the overbar denoting complex conjugation. 
Here, $A_g$ and $B_{3u}$ are irreps of $D_{2h}$, and the $\pm 1$ sign indicates whether the bilinear is even or odd under glide symmetry. 
From (\ref{eq:bilinears_irreps}), the only quadratic terms allowed in the GL theory are $|\Delta^0_{\bm{Q}}|^2$ and $|\Delta^z_{\bm{Q}}|^2$. 
We also identify the allowed fourth-order terms:
\begin{equation}
    |\Delta^0_{\bm{Q}}|^4,\ 
    |\Delta^z_{\bm{Q}}|^4,\ 
    |\Delta^0_{\bm{Q}}|^2|\Delta^z_{\bm{Q}}|^2,\ 
    (\overline {\Delta^0}_{\bm{Q}}\Delta^z_{\bm{Q}} \pm \Delta^0_{\bm{Q}} \overline {\Delta}^{z}_{\bm{Q}})^2.
\end{equation}

Thus, to fourth order, the GL theory takes the form:
\begin{eqnarray}
    S &&= \left(\frac{4}{g} + a_0\right) |\Delta^0_{\bm{Q}}|^2 + \left(\frac{4}{g} + a_z\right) |\Delta^z_{\bm{Q}}|^2 \nonumber\\
&& + b_0
|\Delta^0_{\bm{Q}}|^4 + b_z|\Delta^z_{\bm{Q}}|^4  \nonumber \\
&&+c'\left(
 \bar\Delta^0_{\bm{Q}} \Delta^z_{\bm{Q}} +
 \bar\Delta^z_{\bm{Q}} \Delta^0_{\bm{Q}} \right)^2
 +  c''\left(
 \bar\Delta^0_{\bm{Q}} \Delta^z_{\bm{Q}} -
 \bar\Delta^z_{\bm{Q}} \Delta^0_{\bm{Q}} \right)^2, \nonumber \\
\label{eq:gl_theory_from_symmetry}
\end{eqnarray}
where $g$ is the four-fermion coupling constant defined in (\ref{eq:coulomb_interaction}).
Due to the equivalence of the $\Delta^0_{\bm{Q}}$- and $\Delta^z_{\bm{Q}+\bm{K}_{\bm{a}}}$-phases discussed in the previous section, the coefficients satisfy the relations:
\begin{eqnarray}
    && a_0(\bm{Q}) = a_z(\bm{Q} + \bm{K}_{\bm{a}}), \nonumber \\
    && b_0(\bm{Q}) = b_z(\bm{Q} + \bm{K}_{\bm{a}}).
    \label{eq:gl_theory_coeffs_relations}
\end{eqnarray}
We evaluate these coefficients from the tight-binding model (\ref{eq: 4_4_ham}) with the interaction term in (\ref{eq:coulomb_interaction}); see Appendix \ref{sec:gl_derivation} for a derivation.

To compare to Refs.~\cite{kivelson2006stripes, 2024orbitaltextures}, we first evaluate the coefficients at  $\mu = 1.53$~eV, corresponding to the chemical potential of the undoped tellurium square net. 
The results are shown in Fig.~\ref{fig:a0_az_vs_q_2d}.
The plot of $-a_0(\bm{Q})$ replicates the susceptibility behavior from Ref.~\cite{2024orbitaltextures}, with peaks at approximately $\bm{K}_{\bm{a}} \pm \frac{2}{7} \bm{K}_{\bm{a}}$, $\bm{K}_{\bm{a}} \pm \frac{2}{7} \bm{K}_{\bm{b}}$, related by four-fold rotation symmetry, and their counterparts shifted by multiples of $\bm{K}_{\bm{\alpha}}$ and $\bm{K}_{\bm{\beta}}$.
The plot of $-a_z(\bm{Q})$ can be determined by that of $-a_0(\bm{Q})$ using the relation in Eq.~(\ref{eq:gl_theory_coeffs_relations}).
Fig.~\ref{fig:a0_az_vs_q_2d} shows that if $\mathbf{Q}$ is restricted to the folded BZ, only $\Delta^z_{\bm{Q}}$ exhibits sharp peaks, while $\Delta^0_{\bm{Q}}$ remains relatively smaller and uniform, a trend that we observe for all values of $\mu$.
Thus, henceforth, we allow $\Delta^z_{\bm Q}$ to acquire a nonzero value while $\Delta^0_{\bm Q} = 0$.

In the present work, we are particularly interested in comparing to the experiment on LaSb$_x$Te$_{2-x}$~\cite{bannies2024electronicallydrivenswitchingtopologylasbte}, where we estimate $\mu < 0.4$~eV. 
We study $-a_z(\bm{Q})$ in this regime in Fig.~\ref{fig:az_vs_q_smaller_mu}, which shows that as $\mu$ shrinks, the position of peaks in $-a_z(\bm{Q})$ also shrinks, corresponding to the nesting vector between the two Fermi surfaces in the folded BZ (also shown in Fig.~\ref{fig:az_vs_q_smaller_mu}) shrinking to zero.

\begin{figure}
    \centering
    \includegraphics[height=4.9cm]{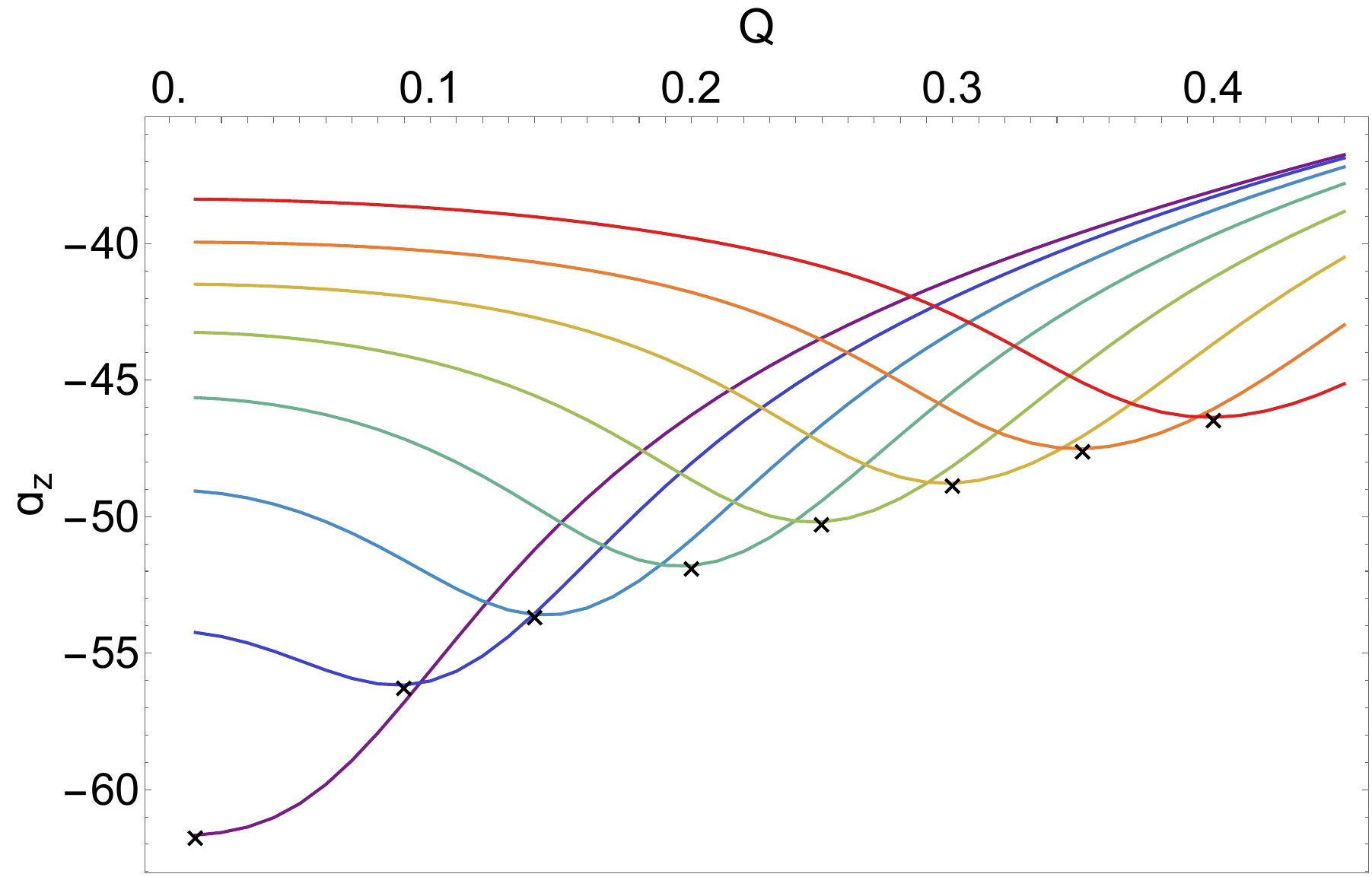}
    \includegraphics[height=4.9cm]{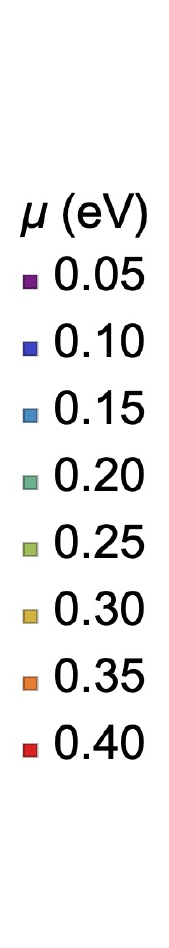}
    \hspace{-1.0cm}
    \vspace{0.4cm} 
    
    \includegraphics[height=4.9cm]{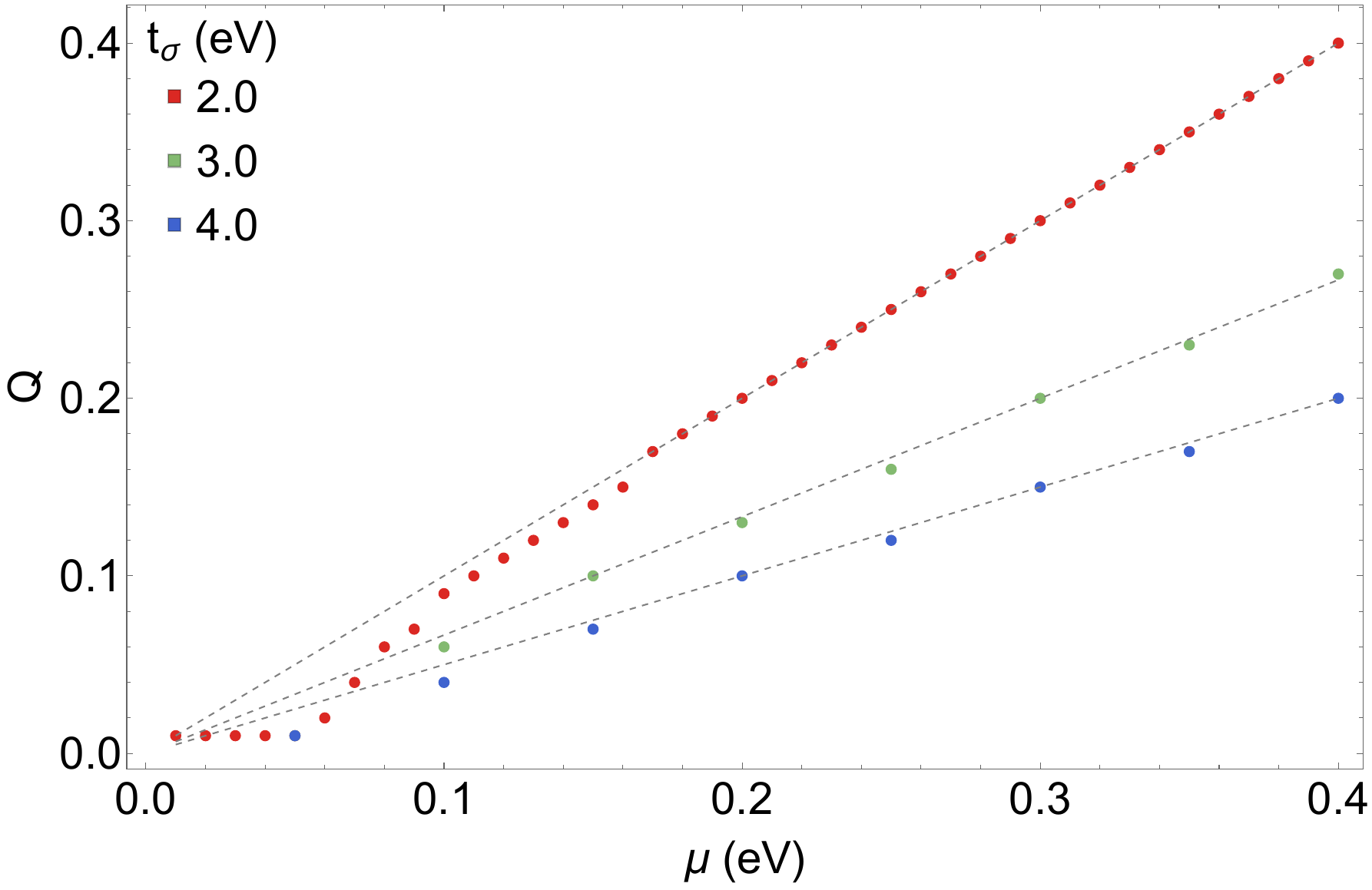}
    \hspace{-0.3cm}
    \vspace{0.2cm} 

    \includegraphics[height=4.9cm]{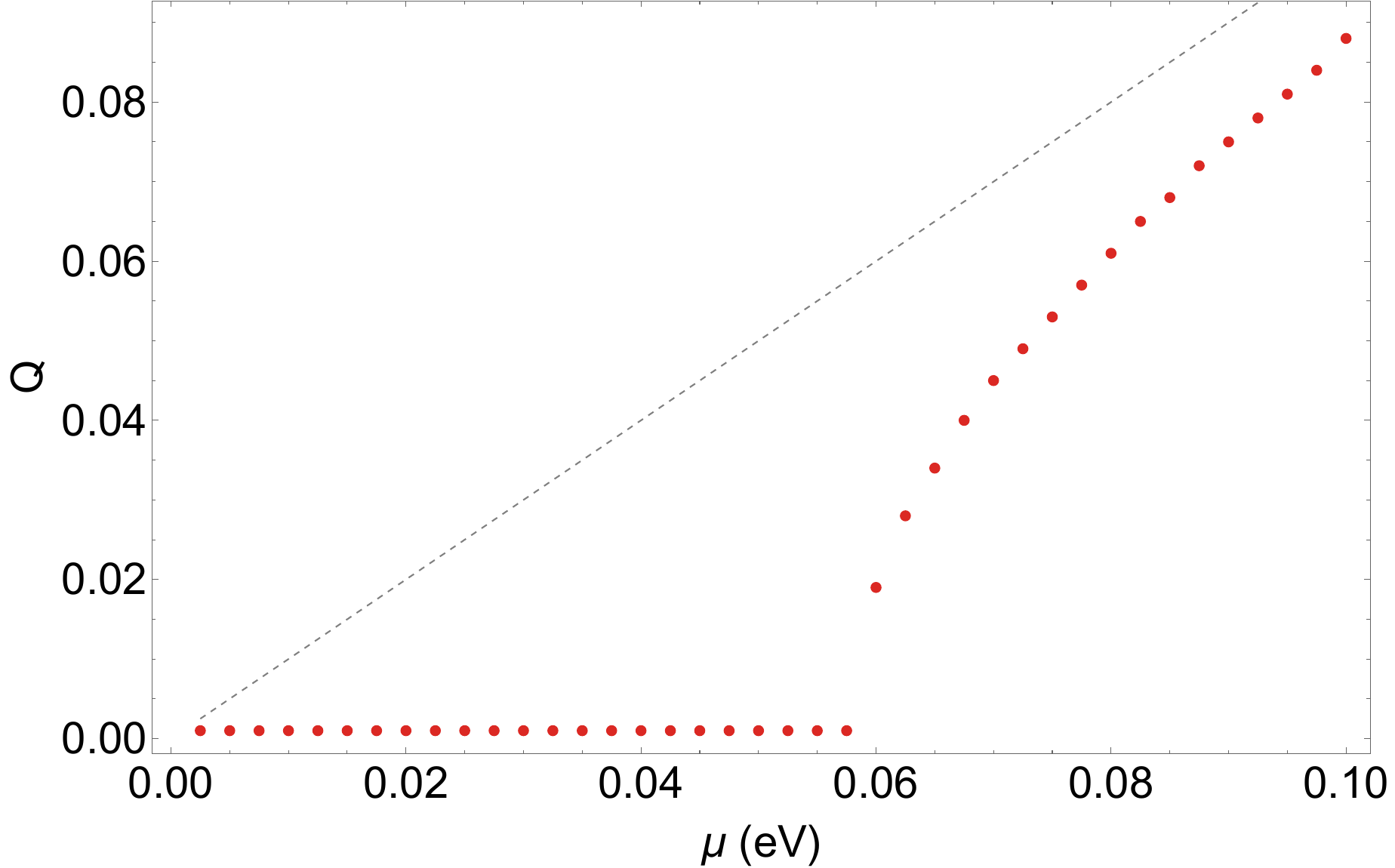}
    \caption{
Upper panel: The quadratic coefficient $a_z(Q, \mu)$ in the GL theory for several fixed values of $\mu$ ranging from $0.05$~eV to $0.40$~eV. Crosses indicate the minima of these functions.
Central panel: $Q$ as a function of $\mu$, obtained by minimizing $a_z(Q, \mu)$ for each fixed value of $\mu$. Grey dashed lines represent $Q=2\mu/t_\sigma$ for three chosen values of $t_\sigma$.
Lower panel:  $Q$ as a function of $\mu$, zoomed in on the region where $\mu$ ranges from $0$ to $0.1$~eV, with $t_\sigma=2.0~{\rm eV}$.
}
    \label{fig:best_q_plots}
\end{figure}

We now study the dependence of $\bm{Q}$ on $\mu$ quantitatively. 
Without loss of generality, we choose $\bm{Q}$ to lie along the $\bm{K}_{\bm{a}}$-direction, i.e. $\bm{Q}=(Q,0)$ in the basis $(\bm{K}_{\bm{a}}/|\bm{K}_{\bm{a}}|,\ \bm{K}_{\bm{b}}/|\bm{K}_{\bm{b}}|)$.
In Fig.~\ref{fig:best_q_plots} (upper panel), we plot $a_z$ as a function of $Q$ for values of $\mu$ ranging from $0.05$~eV to $0.40$~eV.
The lower panels in Fig.~\ref{fig:best_q_plots} display $Q$ as a function of $\mu$, obtained by minimizing $a_z(Q, \mu)$ for each fixed value of $\mu$. 
For generality, we have included multiple values of $t_\sigma$.
Away from $\mu=0$, these plots approximately follow the line $Q = 2\mu/t_\sigma$.
This can be understood because in the limit $t_\sigma \gg t_\pi$, the nesting vector can be approximated by $Q = 4\arcsin\left(\mu/2t_\sigma\right)$~\cite{2024orbitaltextures, kivelson2006stripes}, which approaches $ 2\mu/t_\sigma$ in the limit of small $\mu$.
For small but still finite $\mu$, $Q$ eventually drops to zero. This is a finite-temperature effect, which we discuss at the end of the section.

We now turn to the order parameter $\Delta^z_{\bm{Q}}$.
Since we are studying the regime where $\Delta^0_{\bm{Q}}$ vanishes, we set $\Delta^0_{\bm{Q}}=0$ in Eq.~(\ref{eq:gl_theory_from_symmetry}) to obtain the greatly simplified action:
\begin{equation}
S = \left(\frac{4}{g} + a_z\right) |\Delta^z_{\bm{Q}}|^2
+ b_z|\Delta^z_{\bm{Q}}|^4.
\label{eq:delta_z_action}
\end{equation}
The order parameter is nonzero when $4/g+a_z<0$, taking the value that minimizes Eq.~(\ref{eq:delta_z_action}):
\begin{equation}
    |\Delta^z_{\bm{Q}}|^2 = - \frac{4/g + a_z}{2b_z}.
    \label{eq:gl_minimum}
\end{equation}

\begin{figure}
    \centering
    \includegraphics[height=4.9cm]{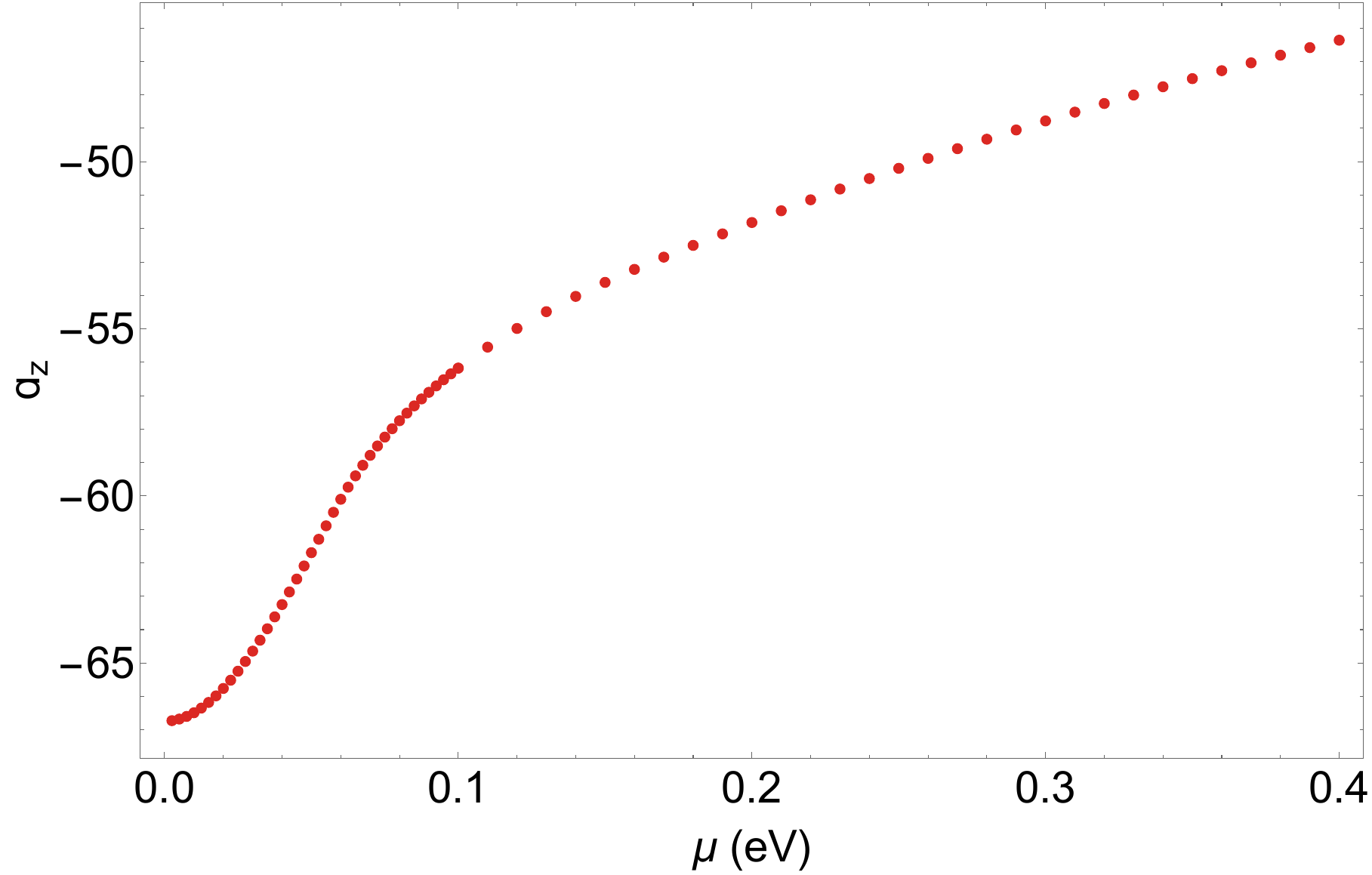}
    \includegraphics[height=4.9cm]{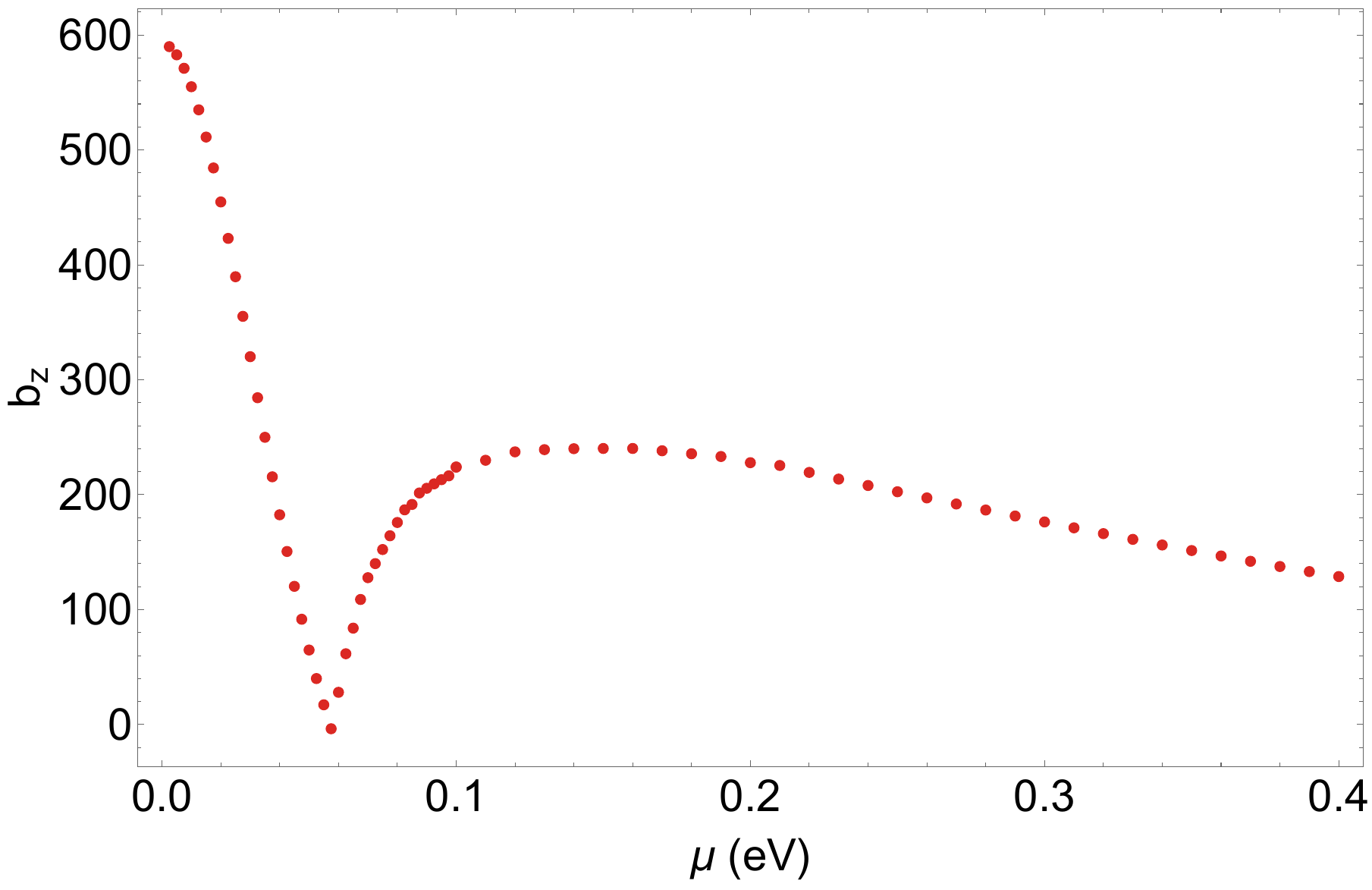}
    \includegraphics[height=4.9cm]{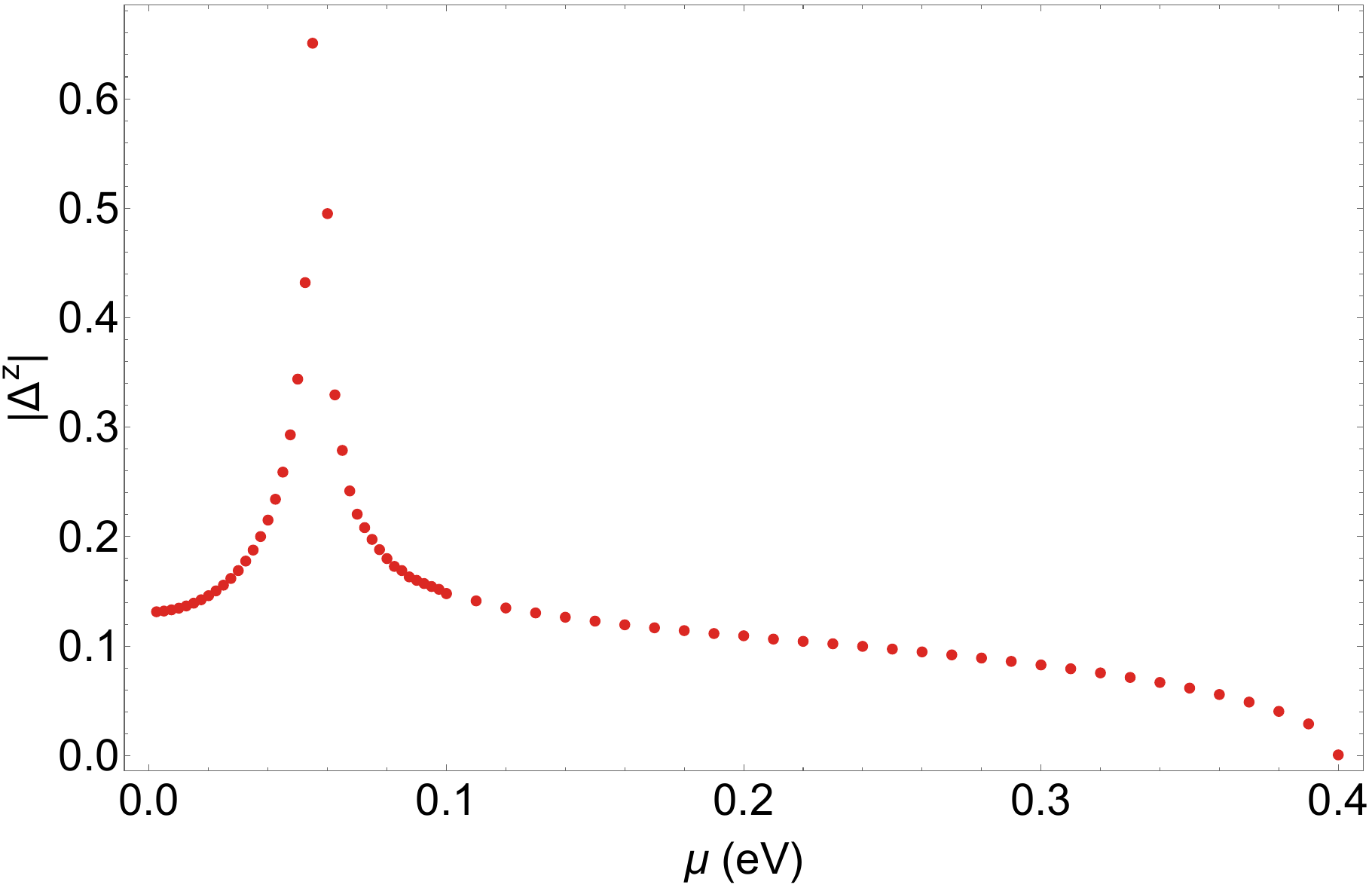}
    \caption{
    Upper panel: The quadratic coefficient $a_z\left(Q(\mu), \mu\right)$ evaluated at the value of $Q$ that minimizes $a_z$ for each $\mu$.
    Central panel: The quartic coefficient $b_z\left(Q(\mu), \mu\right)$ evaluated at the same value of $Q$.
    Lower panel: The order parameter $|\Delta^z_{\bm{Q}(\mu)}|$ that minimizes the free energy. The density of points in the region $\mu < 0.1$~eV is higher to capture the subtle behavior of the coefficient $b_z$.} 
    \label{fig:delta_vs_mu}
\end{figure}

We now study how $\Delta^z_{\bm{Q}}$ depends on $\mu$: for each value of $\mu$, we evaluate $a_z\left(Q, \mu\right)$ (as shown in Fig.~\ref{fig:best_q_plots}) and choose the value of $Q$ that minimizes $a_z$.
The plot of that minimum value of $a_z$ as a function of $\mu$ is shown in Fig.~\ref{fig:delta_vs_mu}, along with the coefficient $b_z\left(Q(\mu), \mu\right)$.
To evaluate the order parameter $\Delta^z_{\bm{Q}}$, shown in the lower panel of Fig.~\ref{fig:delta_vs_mu},
we evaluate Eq.~\ref{eq:gl_minimum} with value of the constant $g$ chosen such that the CDW phase transition (where $\Delta^z_{\bm{Q}} = 0$) occurs at $\mu = 0.4$~eV.
This value of $\mu$ is chosen to approximately match the experimentally observed transition in Ref.~\cite{bannies2024electronicallydrivenswitchingtopologylasbte}.

\begin{figure*}
    \centering
    \includegraphics[height=5cm]{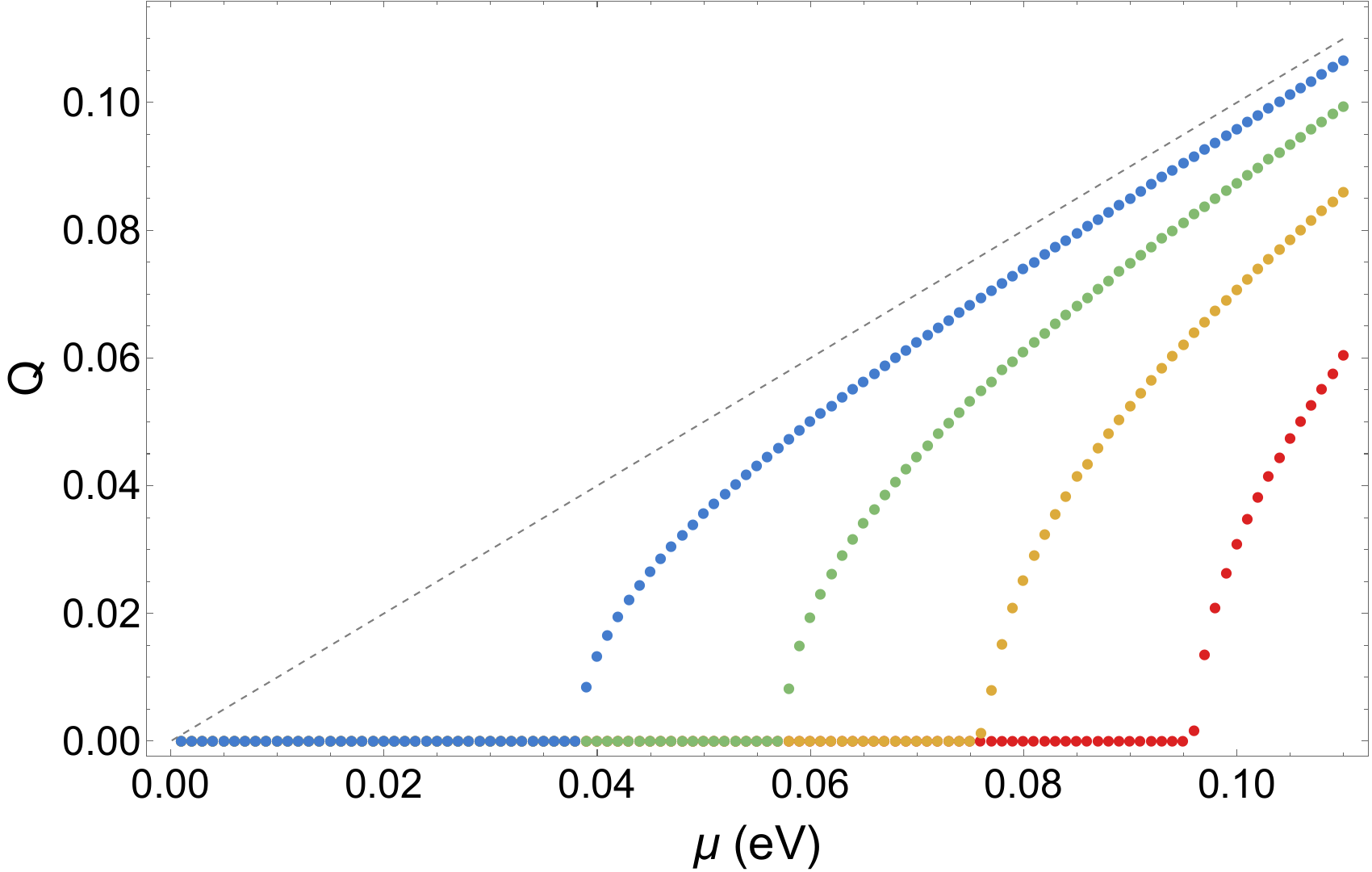}\hspace{0.3cm}
    \includegraphics[height=5cm]{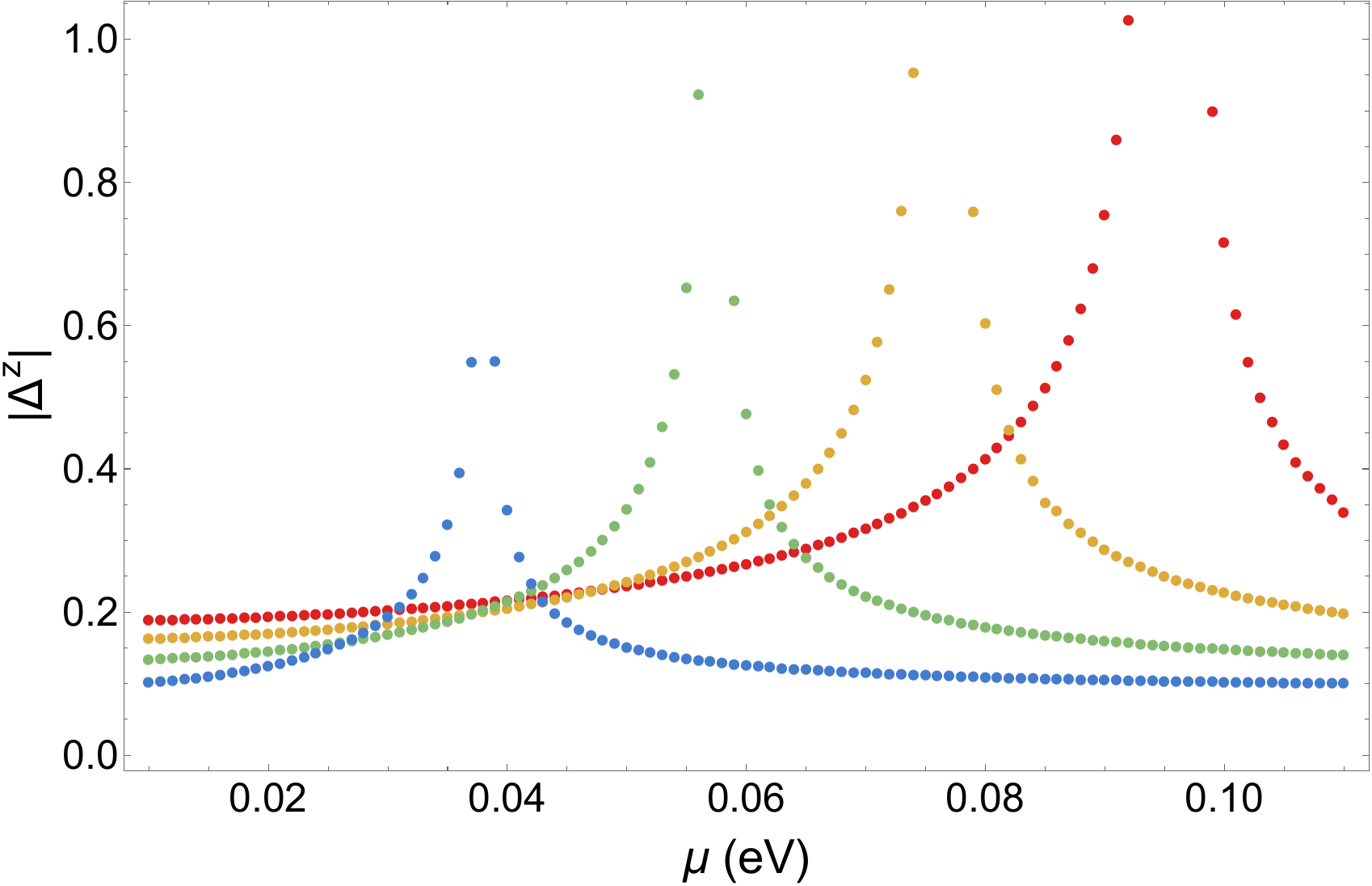}
    \includegraphics[height=5cm]{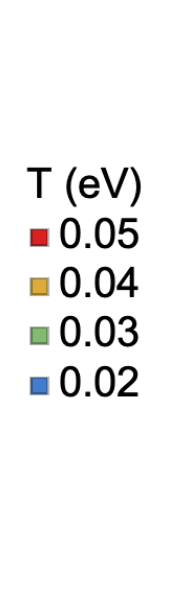}
    \caption{
    Left: $Q$ as a function of $\mu$ for several values of $T$. The grey dashed line indicates $Q = 2\mu / t_\sigma$, with $t_\sigma = 2.0$~eV.
    Right: The order parameter $|\Delta^z_{\bm{Q}(\mu)}|$, minimizing the free energy, for the same set of temperatures; the coupling constant $g$ is fixed across all temperatures and matches the value used in Fig. \ref{fig:delta_vs_mu}.
    }
    \label{fig:T_dependence}
\end{figure*}

The order parameter is finite at $\mu = 0$, has a sharp peak at $\mu \approx 0.06$~eV and then decays monotonically until it reaches zero at $\mu = 0.4$~eV. 
The sharp peak occurs at the value of the chemical potential where $b_z$ (the coefficient of the quartic term in the action (\ref{eq:gl_minimum})) vanishes, which matches the value of chemical potential where $Q$ transitions from a finite value to zero.
We now study the latter as a function of temperature: 
Fig. \ref{fig:T_dependence} shows that the value of the chemical potential at which $Q$ vanishes depends approximately linearly on $T$, and that the peak in $|\Delta^z|$ occurs at that same value of $\mu$ over a range of temperatures.

\section{\label{sec:mft_theory} MFT and gap opening}

\begin{figure*}
    \centering
    \includegraphics[height=4.9cm]{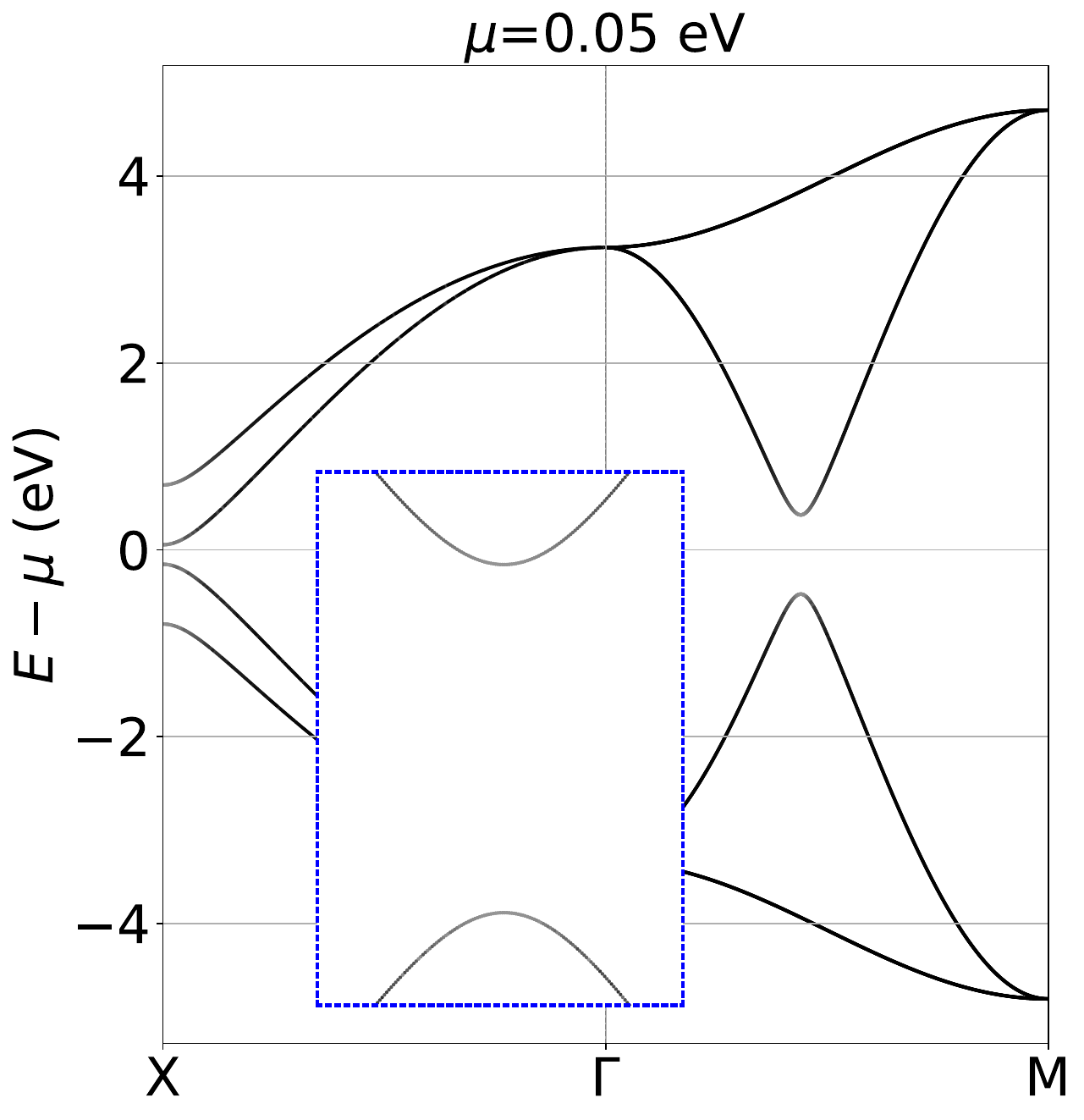}
    \includegraphics[height=4.9cm]{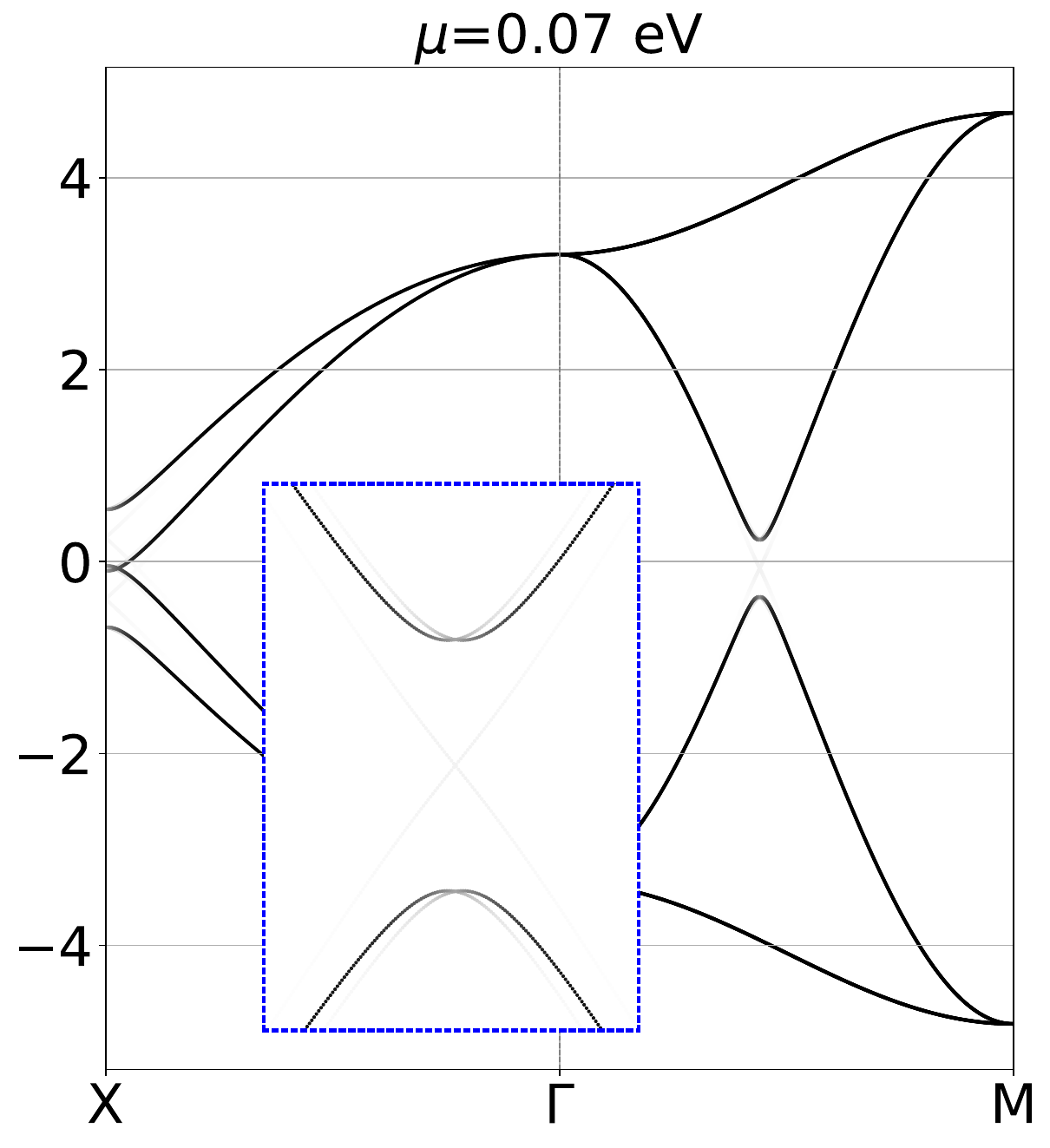}
    \includegraphics[height=4.9cm]{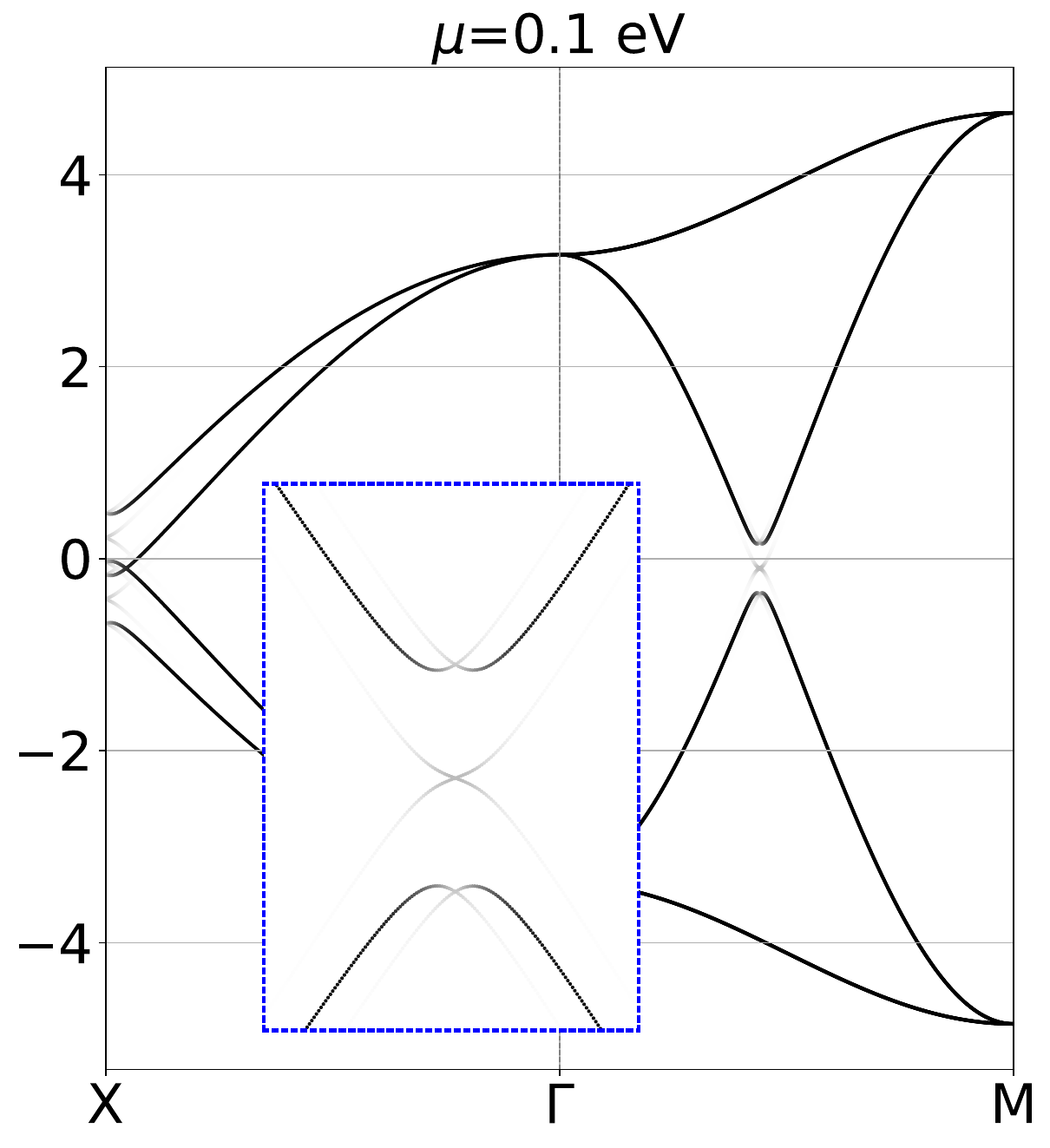}
    \includegraphics[height=4.9cm]{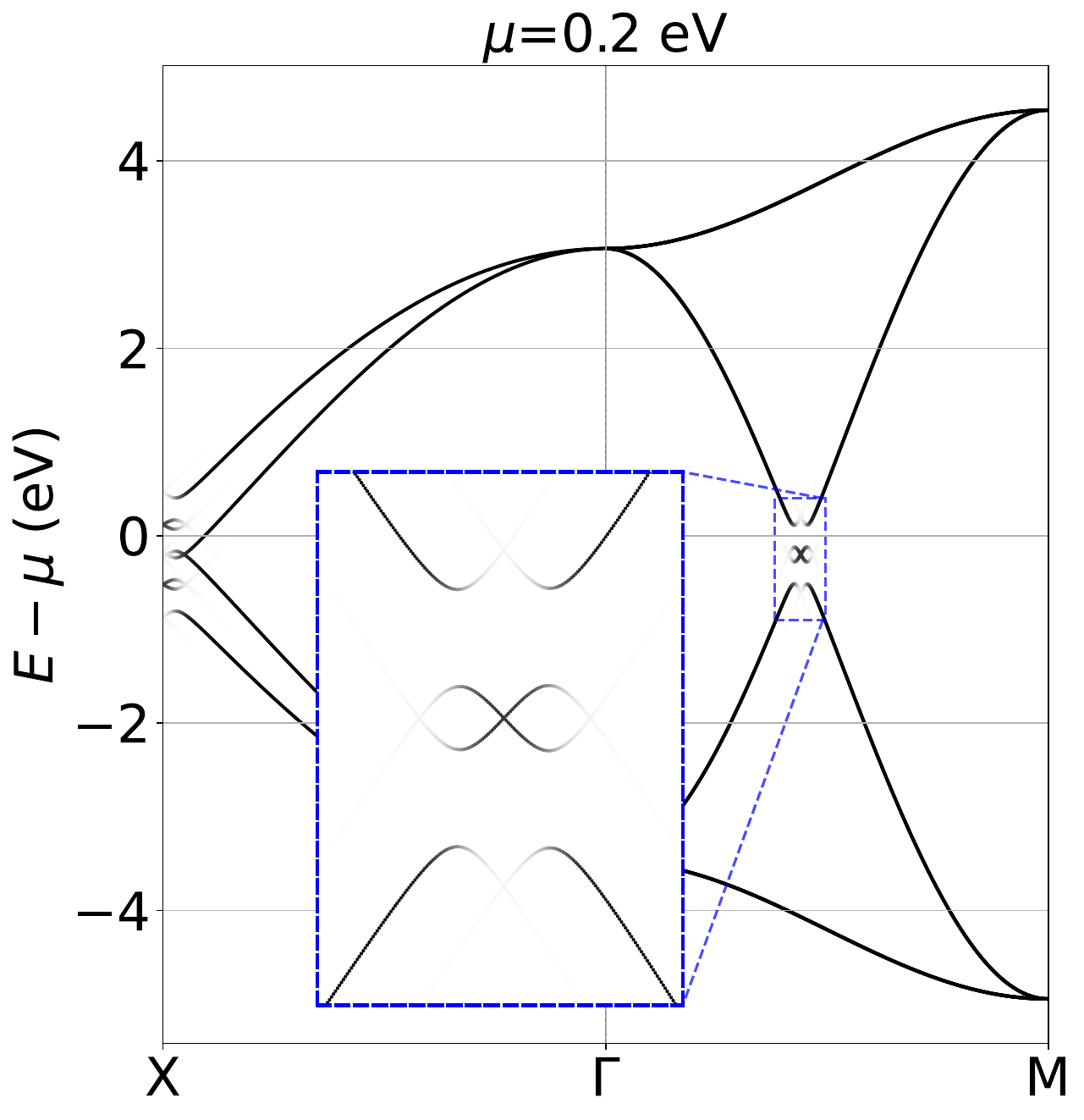}
    \includegraphics[height=4.9cm]{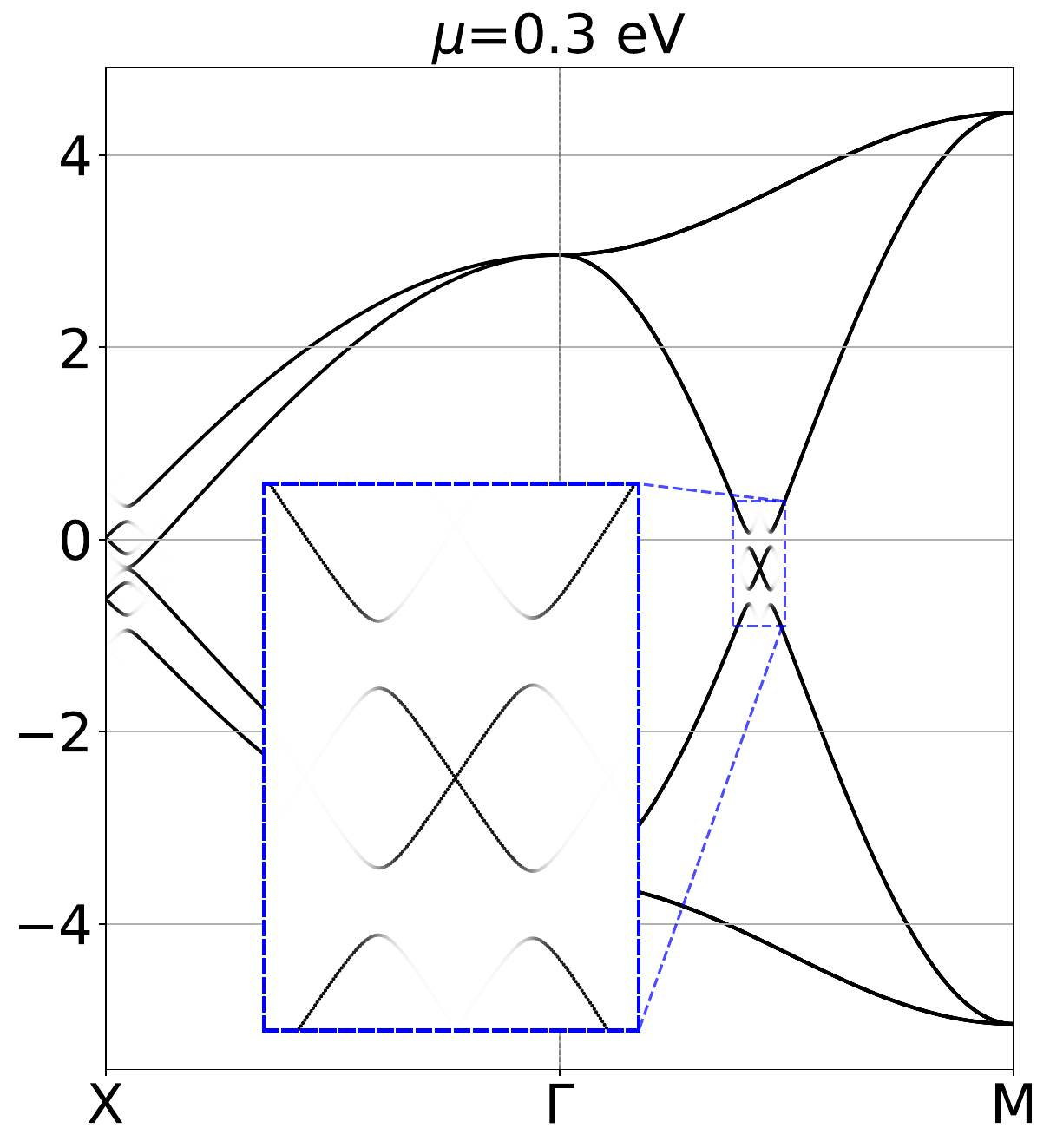}
    \includegraphics[height=4.9cm]{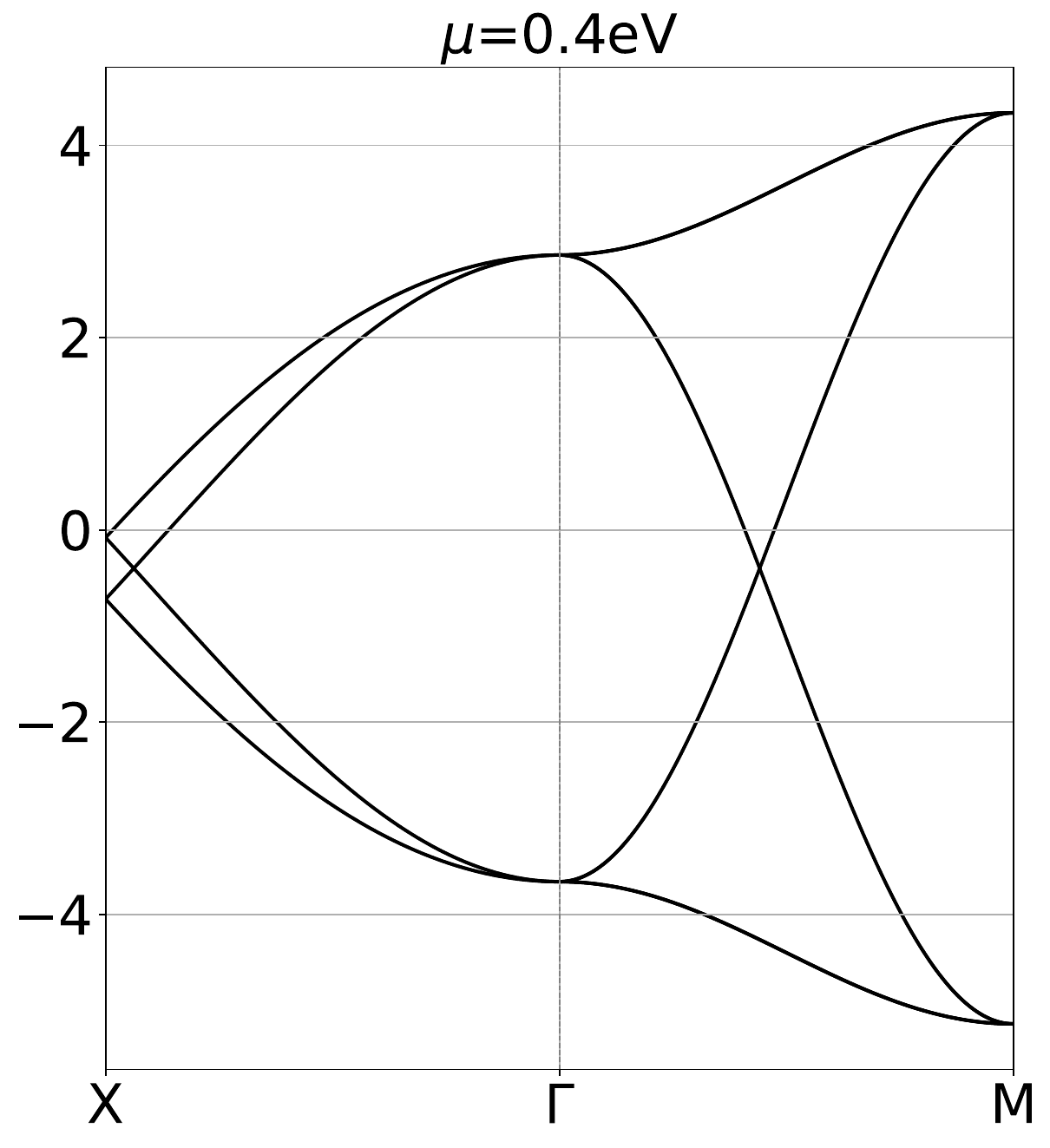}
    
    \includegraphics[height=0.8cm]{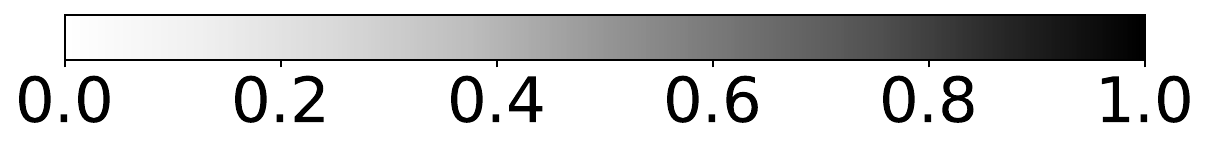}
    \caption{
Energy bands in the presence of the CDW for multiple values of the chemical potential, at fixed temperature $T=0.03$~eV. As $\mu$ decreases, $Q$ decreases as well, while $\Delta^z$ increases, causing the gaps above and below the nodal line to expand. At lower values of 
$\mu$, the in-gap states diminish and eventually disappear. The triplets $(\mu, Q, \Delta^z)$ are as follows: $(0.05, 0, 0.3)$, 
$(0.07, 0.045, 0.2)$, 
$(0.1, 0.085, 0.15)$,
$(0.2, 0.19, 0.12)$,
$(0.3, 0.3, 0.08)$, $(0.4, 0.4, 0)$.
    } 
    \label{fig:energy_bands}
\end{figure*}

We now examine how the order parameters discussed in the previous section can gap the nodal line.
We consider the mean-field Hamiltonian in the presence of a CDW:
\begin{equation}
H_{\operatorname{MFT}} = 
\sum_{\bm{k}}\left[2\psi^\dagger_{\bm{k}}h_{\bm{k}} \psi_{\bm{k}} + \psi^\dagger_{\bm{k}+\bm{Q}}\,
\Delta_{\bm{Q}} \psi_{\bm{k}} + 
\psi^\dagger_{\bm{k}-\bm{Q}}
\Delta_{\bm{Q}}^\dagger \psi_{\bm{k}}\right].
\label{eq:ham_cdw_0}
\end{equation}
Here, $\psi_{\bm{k}}$ is a column vector with components $(\psi_{\bm{k},p_x,A}, \psi_{\bm{k},p_x,B}, \psi_{\bm{k},p_y,A}, \psi_{\bm{k},p_y,B})$ and the CDW-induced term is given by $\Delta_{\bm{Q}} = \Delta_{\bm{Q}}^0 \sigma_0\otimes\tau_0 + \Delta_{\bm{Q}}^z \sigma_0\otimes\tau_z$.
The factor of 2 in the first term in Eq.~(\ref{eq:ham_cdw_0}) follows from the normalization chosen in Eq. (\ref{eq:coulomb_interaction}). This Hamiltonian mirrors the effective action derived in Appendix~\ref{sec:gl_derivation} using a Hubbard-Stratonovich transformation (see Eq. (\ref{eq:mft_like})).
We restrict our attention to the leading instability associated with the nesting vector $\bm{Q}$ and do not include higher-order CDW harmonics such as $\Delta_{2\bm{Q}}$. 
While such terms can renormalize the quartic GL coefficients, their effect is suppressed, as discussed in Ref. \cite{kivelson2006stripes}.
We further truncate Eq.~(\ref{eq:ham_cdw_0}) as follows, which is valid in the limit of small $\Delta_{\bm Q}$:
\begin{equation}
\mathcal{H}_{\bm{k}} = \begin{pmatrix}
h_{\bm{k}-\bm{Q}} & 0 & 0 \\
 0 &   h_{\bm{k}} & 0 \\
 0 &  0 & h_{\bm{k}+\bm{Q}}
\end{pmatrix}
+
\begin{pmatrix}
0 & \Delta_{\bm{Q}}^\dagger & 0 \\
\Delta_{\bm{Q}} & 0 & \Delta_{\bm{Q}}^\dagger \\
0 &  \Delta_{\bm{Q}} & 0
\end{pmatrix}.
\label{eq:ham_cdw}
\end{equation}

In Fig. \ref{fig:energy_bands}, we plot the energy spectrum of $\mathcal{H}_{\bm{k}}$ for several values of the chemical potential $\mu<0.4$~eV, using the values of $\bm{Q}(\mu)$ and $|\Delta^z_{\bm{Q}(\mu)}|$ obtained in the previous section. The color intensity represents the spectral weight in the first BZ, i.e., the sum of the squared magnitudes of the middle four components of each 12-dimensional eigenvector of $\mathcal{H}_{\bm{k}}$.
As $\mu$ and $\bm{Q}$ decrease, $|\Delta^z_{\bm{Q}}|$ increases, leading to the initial appearance of small gap openings below and above the nodal line. These gaps gradually widen as $\mu$ is further reduced. At even smaller values of $\mu$, where $\bm{Q}$ approaches $\bm{0}$, the in-gap states become less pronounced and eventually become almost invisible. 
This behavior results from a redistribution of spectral weight from the in-gap states to the upper and lower bands. In Appendix \ref{sec:in_gap_states}, we introduce a simple toy model that effectively captures the behavior of these in-gap states, demonstrating how a pair of gaps merges into what appears to be a single gap.

We now show that the $\Delta^z$ order parameter is essential for the CDW to open a gap, which is why it is energetically preferred over $\Delta^0$.
First consider the Hamiltonian before the onset of the CDW. For each eigenstate $(\alpha_{\bm{k}}, \beta_{\bm{k}})$ of the unfolded $2\times 2$ Hamiltonian $\tilde{h}_{\bm{k}}$ (\ref{eq: 2_2_ham}) with eigenenergy $\varepsilon_{\bm{k}}$, there exist two corresponding eigenstates of the folded $4\times 4$ Hamiltonian $h_{\bm{k}}$ (\ref{eq: 4_4_ham}) with eigenenergies $\pm\varepsilon_{\bm{k}}$:
\begin{eqnarray}
&&+\varepsilon_{\bm{k}}:
\   
\begin{pmatrix}
\alpha_{\bm{k}} \\
\beta_{\bm{k}}
\end{pmatrix} 
\otimes
\begin{pmatrix}
1 \\
1
\end{pmatrix}, \nonumber  \\
&&-\varepsilon_{\bm{k}}:\  \begin{pmatrix}
\alpha_{\bm{k}} \\
-\beta_{\bm{k}}
\end{pmatrix} \otimes
\begin{pmatrix}
1 \\
-1
\end{pmatrix}.
\end{eqnarray} 
Here, the first factor of the tensor product lies in the orbital space, while the second lies in the sublattice space. From Eq.~(\ref{eq:glide}), it follows that these components have opposite phases under the glide transformation. Hence, where they cross, they cannot gap; this statement remains true even if longer-range terms are added to the Hamiltonian as long as the glide symmetry is preserved.

These protected crossings are exactly what form the nodal loop: along any line $k_y = a k_x$ in momentum space, the eigenstates of the band with negative slope are given by $(\alpha_{\bm{k}}, \alpha_{\bm{k}}, \beta_{\bm{k}}, \beta_{\bm{k}})$, while those of the 
band with positive slope are given by $(\alpha_{\bm{k}}, -\alpha_{\bm{k}}, -\beta_{\bm{k}}, \beta_{\bm{k}})$. 
These bands with opposite slope will cross along the $k_y = ak_x$ line, and, as just discussed, that crossing is protected by the glide symmetry.

\begin{figure}
    \centering
    \includegraphics[height=7cm]{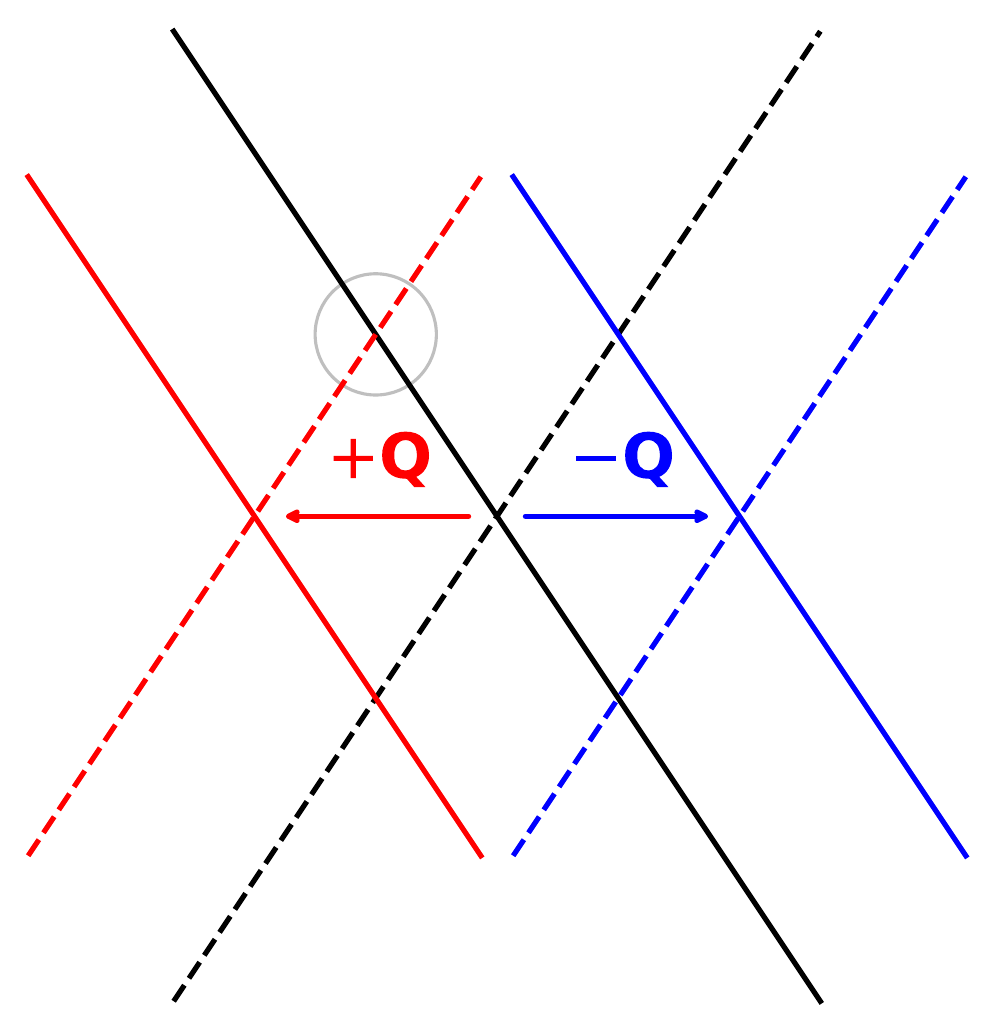}
    \caption{
    Schematic illustration of band crossings between the three Hamiltonians $h_{\bm{k}+
    \bm{Q}}$ (red), $h_{\bm{k}}$ (black), and $h_{\bm{k}-\bm{Q}}$ (blue), for small $\bm{Q}$, along the nodal loop. 
    The bands with negative slope (solid lines) have eigenstates of the form  $(\alpha_{\bm{k}(\pm\bm{Q})}, \alpha_{\bm{k}(\pm\bm{Q})}, \beta_{\bm{k}(\pm\bm{Q})}, \beta_{\bm{k}(\pm\bm{Q})})$, while the bands with positive slope (dashed lines) correspond to eigenstates of the form $(\alpha_{\bm{k}(\pm\bm{Q})}, -\alpha_{\bm{k}(\pm\bm{Q})}, -\beta_{\bm{k}(\pm\bm{Q})}, \beta_{\bm{k}(\pm\bm{Q})})$. At the intersection highlighted by the gray circle, the states involved are $(\alpha_{\bm{k}}, \alpha_{\bm{k}}, \beta_{\bm{k}}, \beta_{\bm{k}})$ and $(\alpha_{\bm{k}+
 \bm{Q}}, -\alpha_{\bm{k}+\bm{Q}}, -\beta_{\bm{k}+\bm{Q}}, \beta_{\bm{k}+\bm{Q}})$. 
 Since these states are orthogonal, they can hybridize only in the presence of $\Delta^z_{\bm{Q}}$ and not with $\Delta^0_{\bm{Q}}$. 
    } 
    \label{fig:band_intersections}
\end{figure}

To describe the CDW, introduce a second set of linearly dispersing bands shifted by a small momentum $\bm{Q}$, corresponding to eigenstates of $h_{\bm{k}+\bm{Q}}$. 
The states from $h_{\bm{k}}$ and $h_{\bm{k}+\bm{Q}}$ with opposite slope may cross, as sketched in Fig.~\ref{fig:band_intersections}.
If the CDW has only a $\Delta^0$ order parameter, the bands that cross cannot hybridize because they are orthogonal. 
However, the $\Delta^z$ component of the CDW order parameter allows for hybridization between these states.
For example, at one of the band intersections between $h_{\bm{k}}$ and $h_{\bm{k}+\bm{Q}}$ highlighted in Fig.~\ref{fig:band_intersections}, the eigenstates of $h_{\bm{k}}$ for the downward-dispersing band take the form $(\alpha_{\bm{k}}, \alpha_{\bm{k}}, \beta_{\bm{k}}, \beta_{\bm{k}})$, while the corresponding states from $h_{\bm{k}+\bm{Q}}$ are $(\alpha_{\bm{k}+\bm{Q}}, -\alpha_{\bm{k}+\bm{Q}}, -\beta_{\bm{k}+\bm{Q}}, \beta_{\bm{k}+\bm{Q}})$. 
Since the overlap between these states vanishes, a $\Delta^0$ order parameter will not allow them to hybridize.
Thus, a gap can only arise from a $\Delta^z$ order parameter.
This can also be understood because the phase factor $-e^{i \bm{\alpha}\cdot \bm{Q}}$ from Eq. (\ref{eq:glide_transformation_phases}), which governs the transformation of the $\Delta^z_{\bm{Q}}$ order parameter under glide symmetry, exactly matches the phase difference between these states: $e^{i \bm{\alpha}\cdot \bm{k}}$ for the former and $e^{-i \bm{\alpha}\cdot (\bm{k}+\bm{Q})}$ for the latter.
This reasoning extends to the other intersections in Fig.~\ref{fig:band_intersections} as well.

To determine whether the band crossing in the CDW state is robust beyond our simplified model, we ask whether there is a symmetry that protects it. Although the CDW breaks the original glide symmetry $\bm{\alpha}$, it may preserve an exact or approximate glide symmetry of the enlarged supercell, depending on the value of $\bm{Q}$. For an exact glide symmetry to exist in the $\Delta^z_{\bm{Q}}$ phase, there must exist integers $n,m$ such that $-e^{i\bm{Q}\cdot(n\bm{\alpha}+m\bm{\beta})}=1$ and the vector $n\bm{\alpha}+m\bm{\beta}$ maps sublattice $A$ onto sublattice $B$ (equivalently, $n+m$ must be an odd integer).
If such a glide symmetry exists, it will protect the nodal line crossing, i.e., there will always be a nonzero density of states at the nodal line, although it may appear vanishingly small within experimental resolution. However, even if an exact glide symmetry does not exist, an approximate one may still be present if the phase condition is approximately satisfied for some integers $n$ and $m$. Such an approximate symmetry may still protect the band crossing within experimental precision; indeed, our minimal model in Eq.~(\ref{eq:ham_cdw}) is agnostic to whether or not there exists an exact glide symmetry, i.e., even if there is no exact glide symmetry for the particular $\mathbf{Q}$, there is always a nonzero density of states at the nodal line crossing when $\mathbf{Q}$ is nonzero, as is apparent from Fig.~\ref{fig:energy_bands}. Only for the smallest value of chemical potential in that figure, $\mu = 0.05$~eV, does the nodal line completely disappear, because for this value of $\mu$, $\mathbf{Q}$ precisely vanishes.

The $\mathbf{Q} =0$ case admits a simple physical interpretation: in this case, the order parameter is explicitly odd under the glide symmetry and the Hamiltonian reduces to that in Eq.~(\ref{eq: 4_4_ham}) with an additional $\sigma_0\otimes \tau_z$ term that explicitly breaks glide symmetry. 
In the resulting phase, the electron densities on sublattices $A$ and $B$ become uniform and opposite: $\Delta^A(\bm{R})=-\Delta^{B}(\bm{R})=\text{const}$.
Since the symmetry between the two sublattices is broken,
the nodal line is fully gapped.

\section{\label{sec:experiment}Comparison to experiment}
We now discuss the experimental results.
Ref.~\cite{bannies2024electronicallydrivenswitchingtopologylasbte} demonstrates that electron doping can tunably gap and restore the nodal line in LaSb$_x$Te$_{2-x}$.
Specifically, the study compares two methods of changing the electron filling in the square-net layer: chemical substitution of Te with Sb and potassium (K) deposition/desorption. In both cases, the authors report the opening of a nodal-line gap. 
In the first scenario, increasing the Sb concentration
drives a structural phase transition from an orthorhombic $(x=0.7)$ to tetragonal $(x\approx 0.85)$ phase and back again to an orthorhombic phase $(x=1)$. When $x<0.9$, the nodal line is visible in ARPES, while when $x>0.9$, it appears gapped.

In the K deposition experiment, the system is initially in the gapped phase with $x>0.9$. As the K concentration is increased, the nodal line gap closes. This is a reversable process: desorption of potassium restores the gapped state.

Our theory captures the trends in both experiments (though we do not describe the orthorhombic phase when $x=0.7$).
Specifically, the increase in Sb concentration from $x\approx 0.85$ to $x=1$ corresponds to lowering the chemical potential in our model, which we have shown (e.g., in Fig.~\ref{fig:energy_bands}) induces a nodal-line gap through the formation of a CDW.

In the potassium experiment, increasing the K concentration corresponds to increasing the chemical potential $\mu$ in our model. Based on our estimates, the experiment begins at $\mu \approx 0.1$~eV — where the nodal line is gapped — and ends around $\mu \approx 0.4$~eV, where the gap is fully closed. 
Thus, this experiment is also consistent with our model.
Since the experimental temperature ($T = 20$K) is lower than the temperature used in our calculations, we expect the experiment lies in the regime where the CDW ordering vector $\bm{Q}$ remains nonzero, based on our estimated hopping amplitudes. However, we cannot rule out the possibility that $\bm{Q} = 0$ in the experiment, as different hopping parameters could shift the critical value of $\mu$ at which $\bm{Q}$ vanishes.

In summary, the evolution of the nodal-line gap observed in the experiment closely parallels our theoretical results shown in Fig.~\ref{fig:energy_bands}: in-gap states emerge within the nodal loop gap at intermediate doping, and further doping eventually leads to complete gap closure.

\section{\label{sec:conclusion} Conclusion}

Our work resolves open questions about the interplay between CDW order and nodal line topology. 
Specifically, our study elucidates the mechanisms by which the CDW order can affect symmetry-protected nodal lines in materials such as LaSb$_x$Te$_{2-x}$.
While a CDW with finite wave vector $\mathbf{Q}$ need not break the glide symmetry that protects the nodal line, it can nonetheless significantly suppress the spectral weight at the crossing — particularly when the chemical potential approaches the nodal line. 
When $\mathbf{Q} = 0$, the resulting order fully breaks the glide symmetry and a true gap opens at the nodal point.

Even for a finite wave vector, if the chemical potential is sufficiently close to the nodal line, the suppression of spectral weight at the nodal line crossing will result in an apparently gapped nodal line in, e.g., an ARPES experiment.
Thus, the evolution and disappearance of in-gap states predicted by our theory is consistent with the experimentally observed behavior in LaSb$_x$Te$_{2-x}$ \cite{bannies2024electronicallydrivenswitchingtopologylasbte}, as well as earlier work on GdSb$_x$Te$_{2-x}$ \cite{lei2021band}.

At finite temperature, if the chemical potential is nearby but not exactly at the nodal line, we identify a $\mathbf{Q} = 0$ instability that breaks the glide symmetry and gaps the nodal line.
Our calculations indicate that the critical value of chemical potential will increase linearly with temperature.

It is an open question whether the orbitally ordered CDW introduced in the context of the related $R$Te$_3$ family \cite{2024orbitaltextures} is also present in the $R$SbTe family. If so, additional symmetry breaking is possible, which may also impact symmetry-protected band crossings. In addition, the interplay between magnetism and conventional and unconventional CDWs in this family of materials remains a rich topic for future work.

\begin{acknowledgments}

S.A. and J.C. are grateful for conversations with Ken Burch, Rafael Fernandes, Sayed Ghorashi, and Leslie Schoop during prior collaborations.
L.C. and J.C. acknowledge support from the Air Force Office of Scientific Research under Grant No. FA9550-24-1-0222.
J.C. acknowledges support from the Flatiron Institute, a division of the Simons Foundation. 
This work was performed in part at the Aspen Center for Physics, which is supported by National Science Foundation grant PHY-2210452. This work was performed in part at the Aspen Center for Physics, which is supported by a grant from the Simons Foundation (1161654, Troyer).

\end{acknowledgments}

\appendix

\section{\label{sec:gl_derivation}Derivation of Ginzburg-Landau theory}
\begin{figure*}
    \centering
    \includegraphics[height=3.9cm]{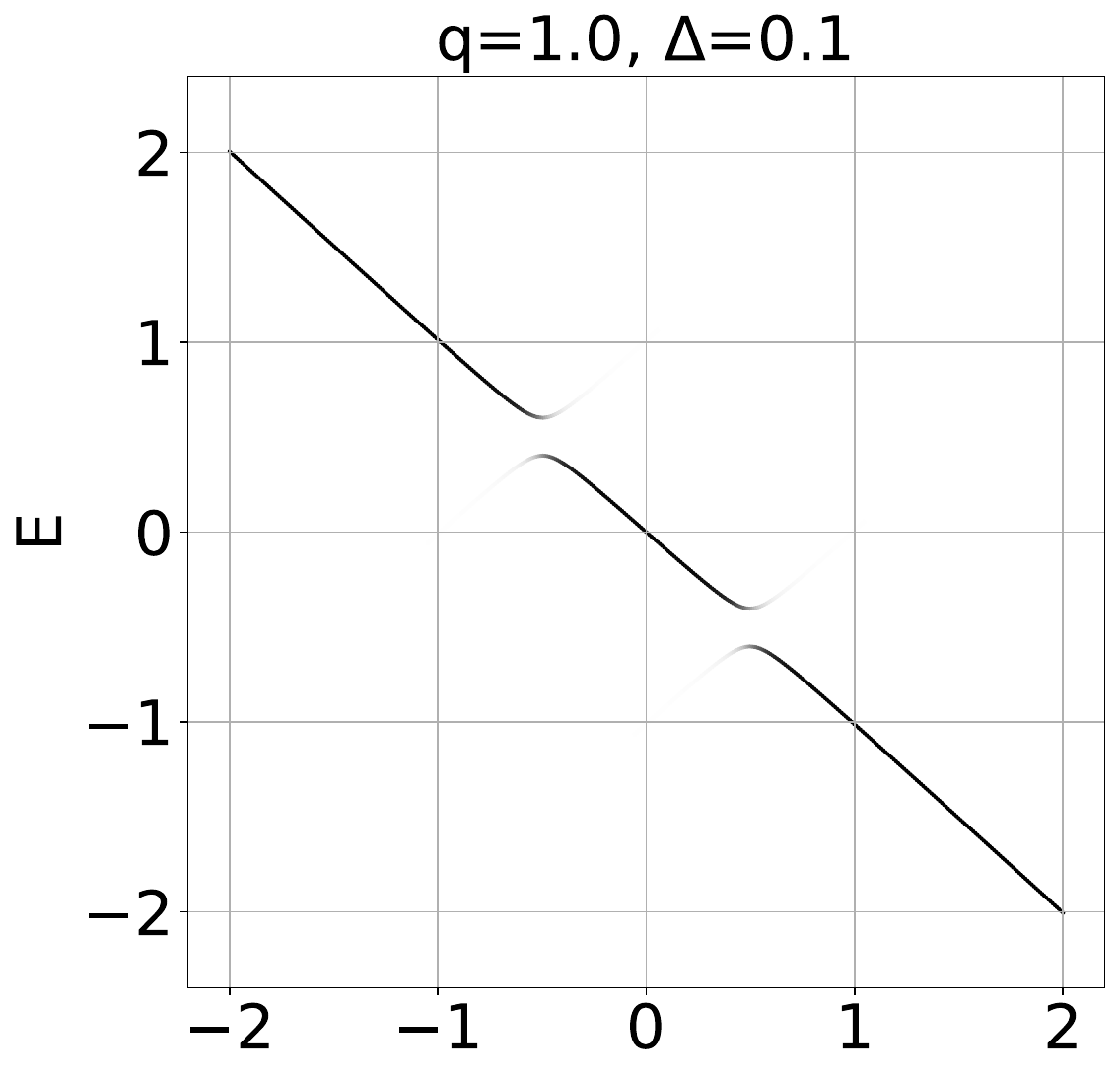}
    \includegraphics[height=3.9cm]{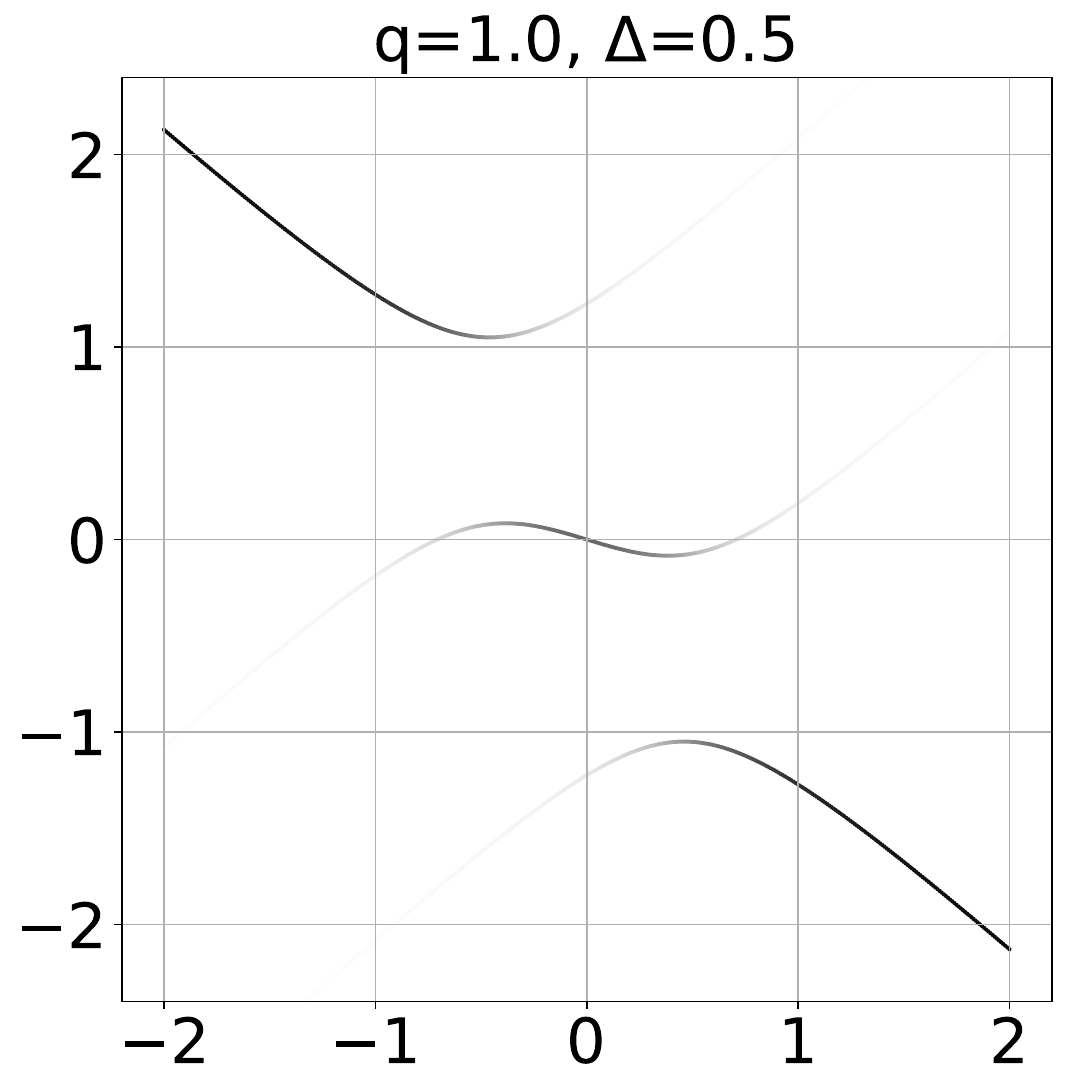}
    \includegraphics[height=3.9cm]{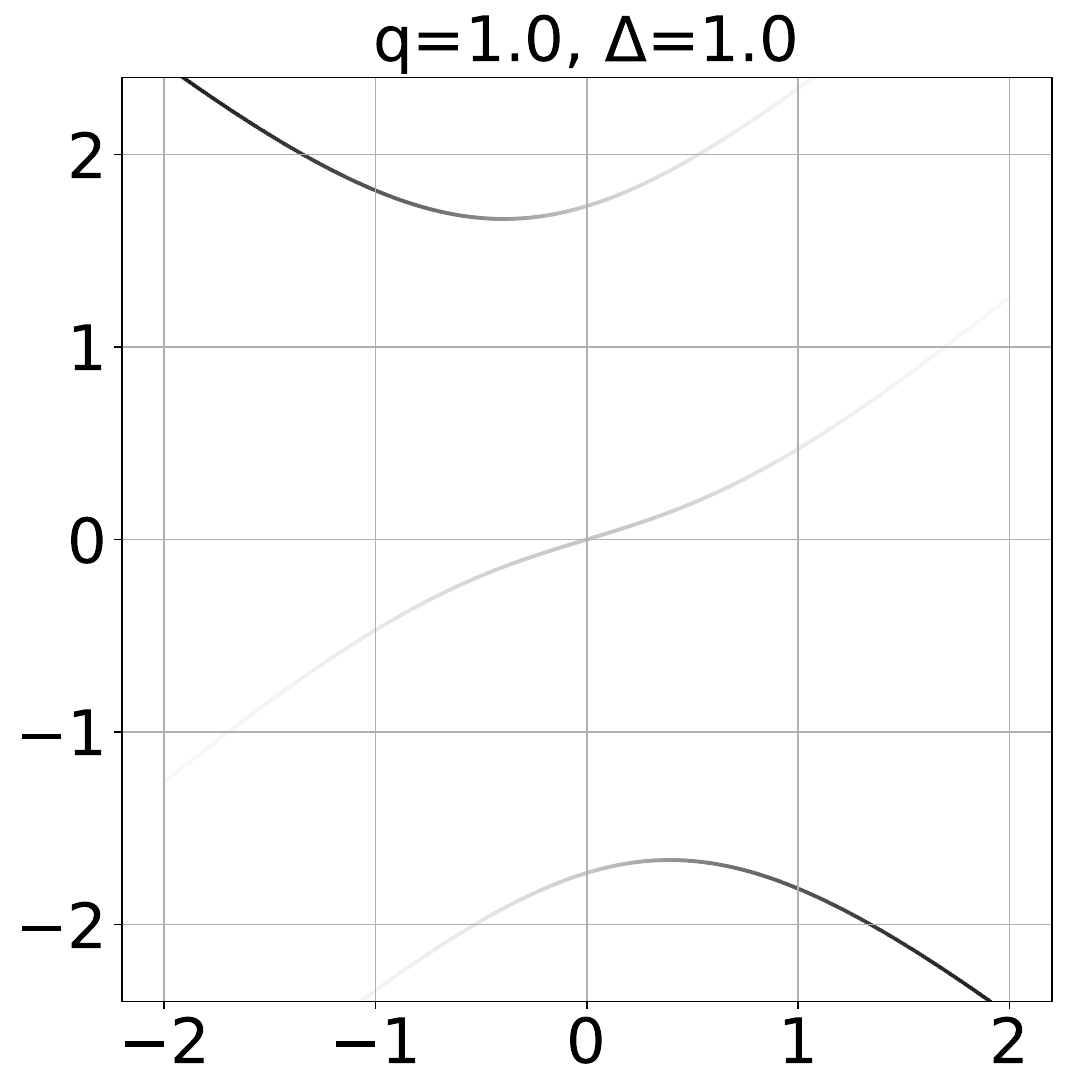}
    
    \includegraphics[height=3.9cm]{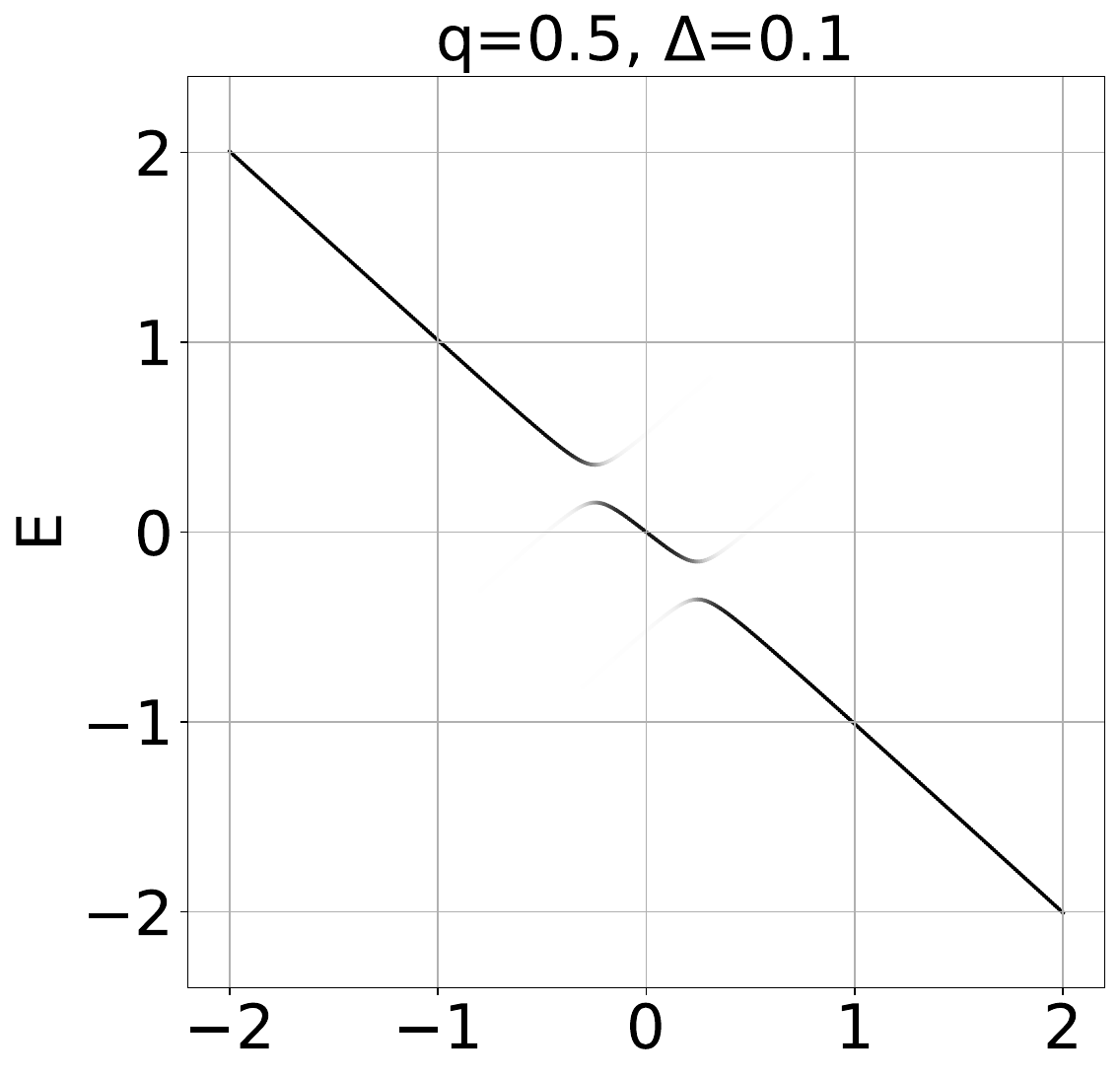}
    \includegraphics[height=3.9cm]{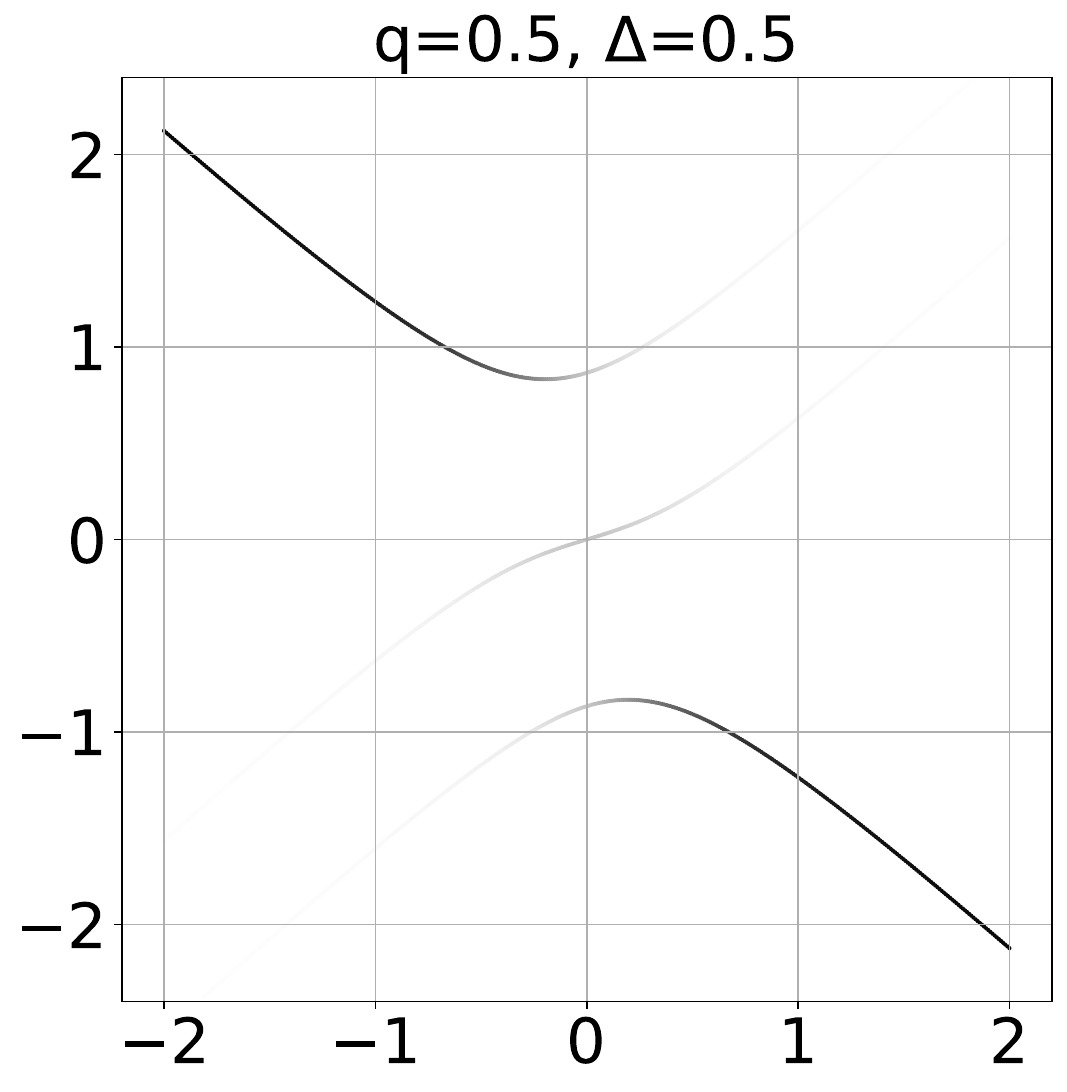}
    \includegraphics[height=3.9cm]{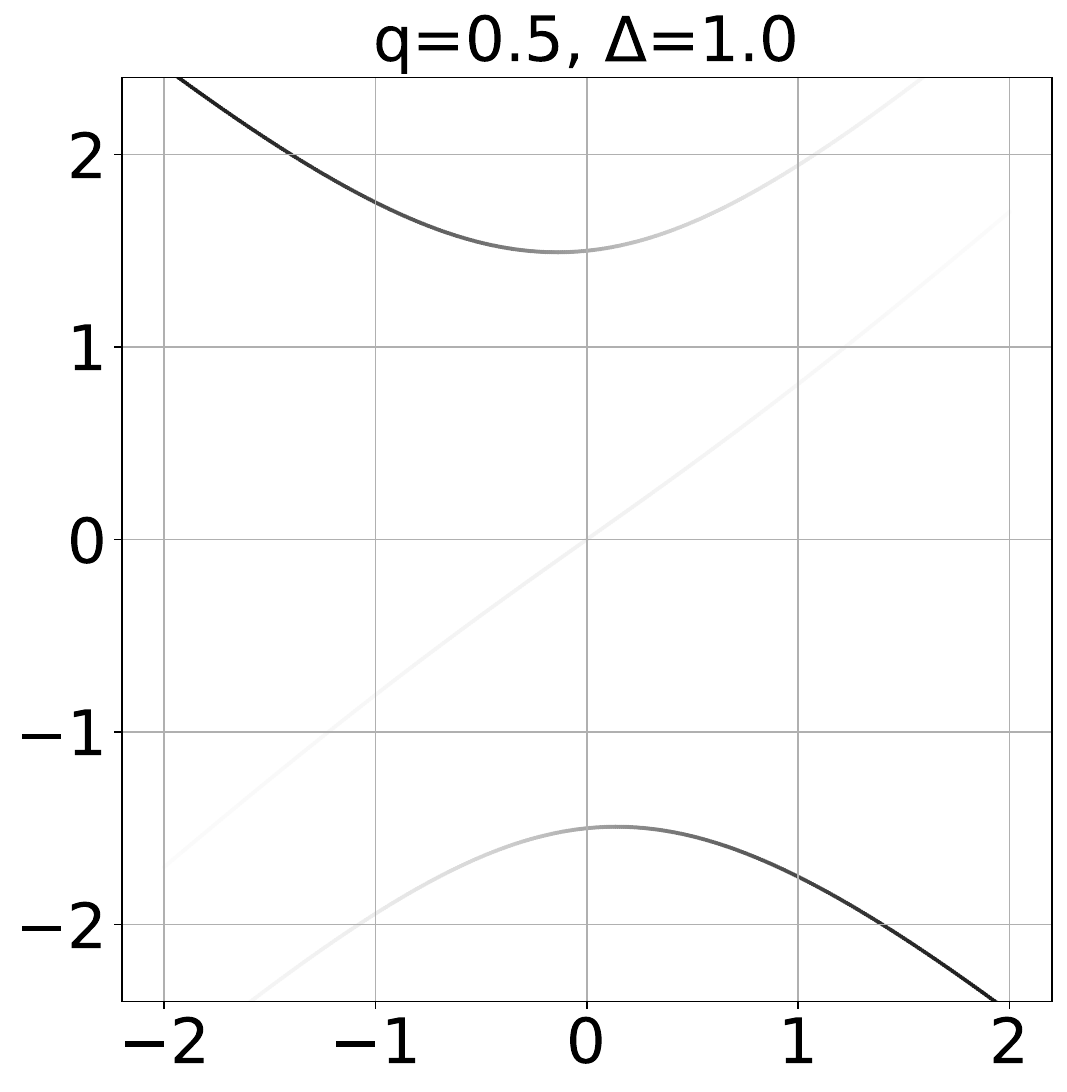}
    
    \includegraphics[height=4.1cm]{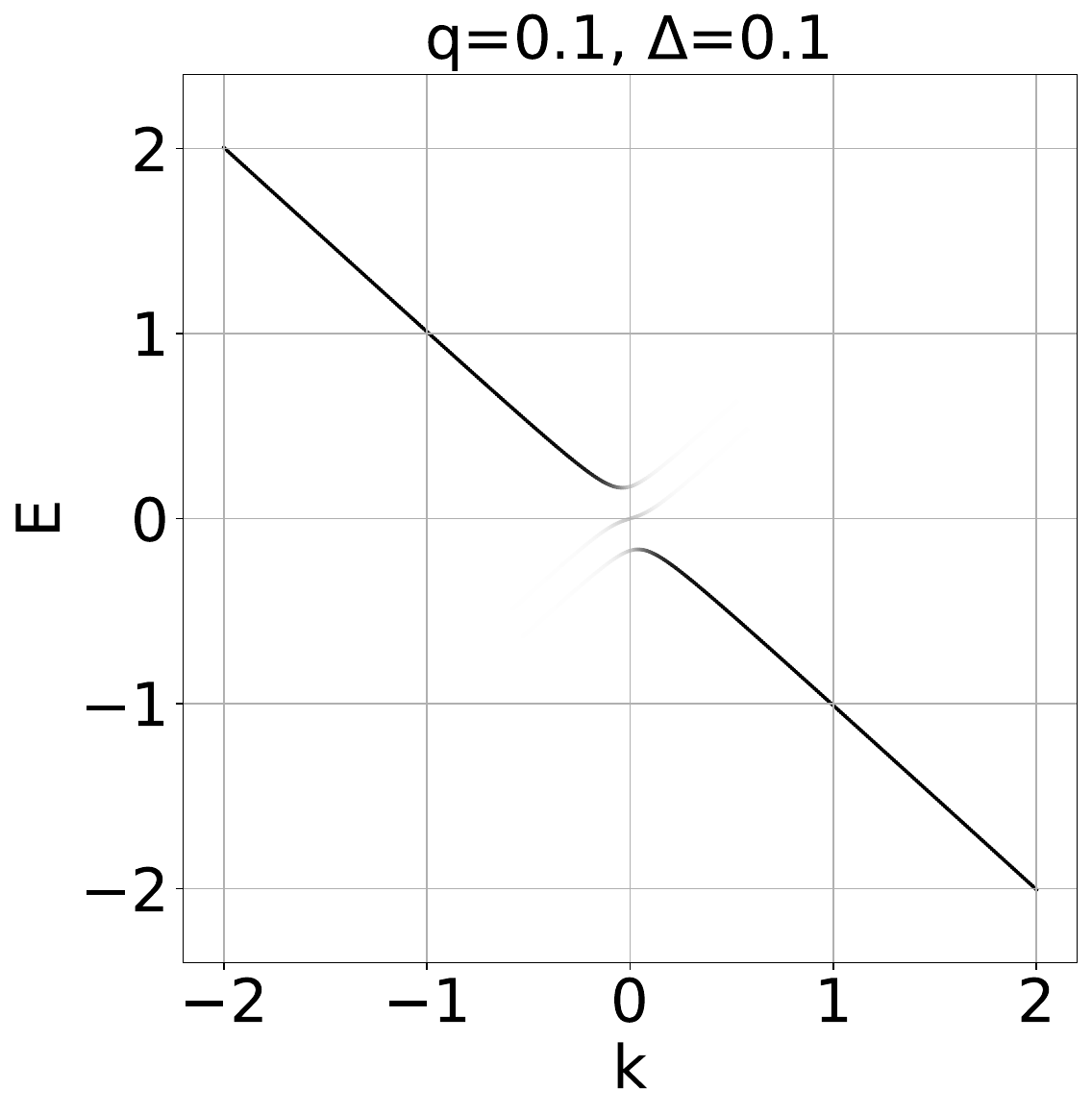}
    \includegraphics[height=4.1cm]{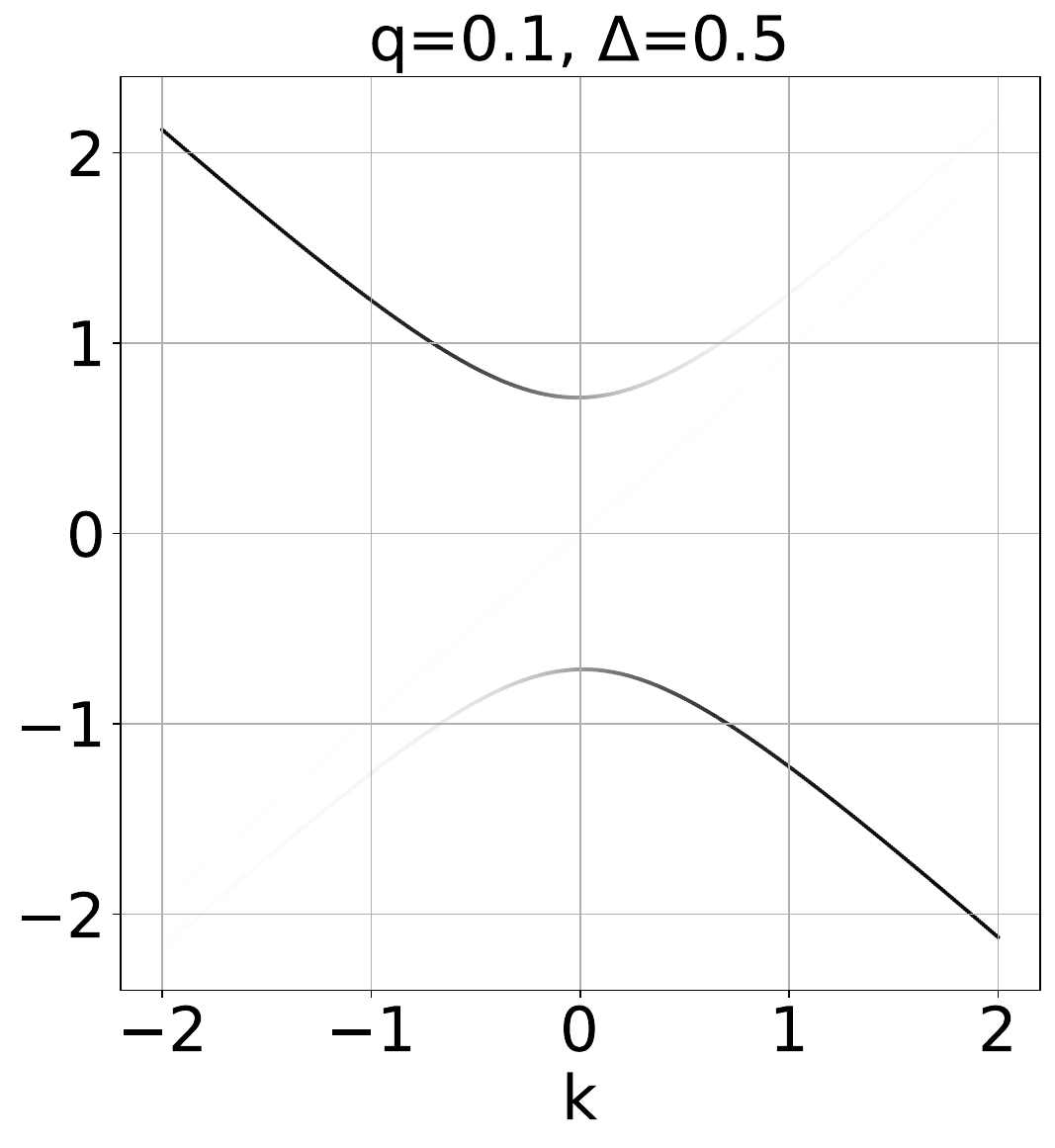}
    \includegraphics[height=4.1cm]{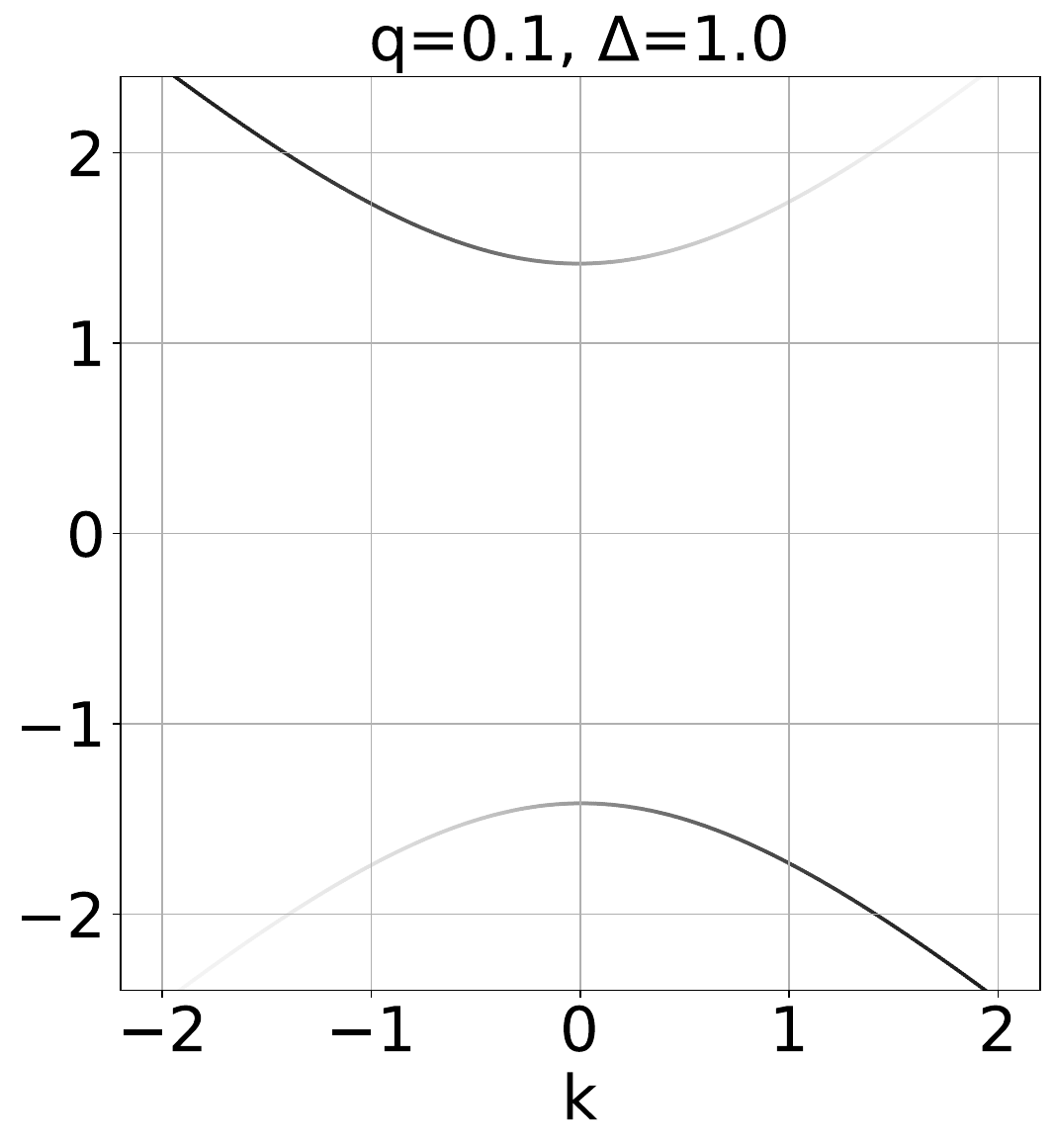}
    
    \includegraphics[height=0.8cm]{graphics/energy_bands/colorbar.pdf}
    \caption{
Energy bands of the simplified Hamiltonian (\ref{eq:toy_ham}) computed for a grid of parameters where $q$ and $\Delta$ take the values $0.1,\, 0.5,\, 1.0$.
    } 
    \label{fig:toy_model}
\end{figure*}
For completeness, we present the derivation of GL theory from the tight-binding model. The derivation closely follows Ref.~\cite{2024orbitaltextures}, with the main difference being the inclusion of sublattice indices in the fermion operators and order parameters.
To obtain the action in terms of the order parameters $\Delta^{0,z}_{\bm{Q}}$, we start with the tight-binding Hamiltonian $h_{\bm{k}}$ (\ref{eq: 4_4_ham}) and the interaction term $V$ (\ref{eq:coulomb_interaction}), where the summation is restricted to a single $\bm{Q}$.
We then express the partition function as an integral over Grassmann variables and decouple the quartic term $V$ using the Hubbard-Stratonovich transformation:
\begin{widetext}
\begin{equation}
Z \propto \int d \bar \Delta d \Delta\, \exp\left(
-\sum_{\alpha I, \beta J} \int d\tau \frac{\bar\Delta_{\bm{Q}}^{\alpha I,\beta J} \Delta_{\bm{Q}}^{\alpha I,\beta J}}{g}
\right) \int d\bar\psi d\psi\, e^{-S}.
\label{eq:HS_transform}
\end{equation}
\end{widetext}
Here, $S$ is the fermionic action, modified by the presence of $\Delta_{\bm{Q}}^{\alpha I,\beta J}$:
\begin{eqnarray}
    &&S = \int d\tau \bigg[2
\sum_{\bm{k}}\psi^\dagger_{\bm{k}}\left(\partial_\tau + h_{\bm{k}} \right)\psi_{\bm{k}}+ 
\sum_{\alpha I, \beta J} \bar A_{\bm{Q}}^{\alpha I, \beta J} \Delta_{\bm{Q}}^{\alpha I, \beta J }  \nonumber \\
&& +\sum_{\alpha I, \beta J} \bar \Delta_{\bm{Q}}^{\alpha I, \beta J} A_{\bm{Q}}^{\alpha I, \beta J}\bigg],
\label{eq:mft_like}
\end{eqnarray}
where $A_{\bm{Q}}^{\alpha I, \beta J}$ denotes a fermionic bilinear of the following form:
\begin{equation}
A_{\bm{Q}}^{\alpha I, \beta J} = \sum_{\bm{k}}\bar\psi_{\bm{k-Q},\beta J}\psi_{\bm{k},\alpha I}.
\end{equation}

In the expressions above, $\alpha I$ is a multi-index composed of the orbital index $\alpha \in \{p_x, p_y\}$ and the sublattice index $I \in\{A,B\}$, while $\psi_{\bm{k}}$ is a column vector with components $\psi_{\bm{k},\alpha I}$.
Although in Eq. (\ref{eq:HS_transform}) we assume that $\Delta_{\bm{Q}}^{\alpha I,\beta J}$ takes the specific form $\sigma_0^{\alpha\beta}\left(\Delta^0_{\bm{Q}}\tau_0^{IJ} + \Delta^z_{\bm{Q}} \tau_z^{IJ}
\right)$, we keep summation over all indices for convenience and for easier comparison with Ref.~\cite{2024orbitaltextures}.
Integrating out the fermions in Eq. (\ref{eq:HS_transform}), one obtains an effective action in terms of the order parameters $\Delta_{\bm{Q}}^{\alpha I, \beta J}$: 
\begin{equation}
    S_{\operatorname{eff}} = 
    \frac{1}{g}
\sum_{\alpha I, \beta J}\bar\Delta_{\bm{Q}}^{\alpha I, \beta J} \Delta_{\bm{Q}}^{\alpha I, \beta J} - 
    \operatorname{Tr} \log \left(1- \mathcal{G} \mathcal{V}\right),
    \label{eq:eff_action_general_d2h}
\end{equation}
where $\mathcal{G}$ is a matrix of non-interacting Green's functions, $\mathcal{V}$ is the interaction matrix, and $\operatorname{Tr}$ denotes summation over momenta, Matsubara frequencies, and matrix indices:
\begin{equation}
    \operatorname{Tr} = \frac{1}{\beta} \sum_{i \omega_n} \int \frac{d \bm{k}}{(2 \pi)^2}\ \operatorname{tr}_{\alpha I,\beta J},
\end{equation}
where $\bm{k}$ runs over the folded Brillouin zone, i.e., $0 \leq k_x, k_y < 2\pi$, and $\omega_n = (2n+1)\pi/\beta$ are fermionic frequencies. To evaluate the full trace, we first sum over the matrix indices, then compute the frequency sums by replacing the summation with a contour integral and evaluating the residues at the poles, and finally perform numerical integration over the momenta.
The matrix of non-interacting Green's functions is given by:
\begin{equation}
    \mathcal{G}_{\bm{k}}(i\omega_n) = \begin{pmatrix}
    G_{\bm{k}}(i\omega_n) & 0 \\
    0 & G_{\bm{k+Q}}(i\omega_n)
    \end{pmatrix},
\end{equation}
where $G_{\bm{k}}(i\omega_n)$, the Green's function of the original system (\ref{eq: 4_4_ham}), is a $4\times4$ matrix. The interaction matrix takes the form:
\begin{equation}
    \mathcal{V} = 
    \begin{pmatrix}
        0 & \Delta_{\bm{Q}}^\dagger \\
        \Delta_{\bm{Q}} & 0 
    \end{pmatrix}
\label{eq: int_matrix}
\end{equation}
with $\Delta_{\bm{Q}} = \Delta_{\bm{Q}}^0 \sigma_0\otimes\tau_0 + \Delta_{\bm{Q}}^z \sigma_0\otimes\tau_z$. Expanding the expression (\ref{eq:eff_action_general_d2h}) in powers of $\Delta^{0,z}_{\bm{Q}}$, we obtain the effective action for these order parameters up to fourth order:
\begin{widetext}
    \begin{eqnarray}
    &&S_{\operatorname{eff}} =  \frac{4}{g} 
    \sum_{i=0,z}
    | \Delta^i_{\bm{Q}} |^2 +\sum_{ij} \operatorname{Tr} \left[ 
    G_{\bm{k}} \,
    \sigma_0 \otimes \tau_i \,
    G_{\bm{k+Q}} \,
    \sigma_0 \otimes \tau_j
    \right] 
    \bar \Delta^i_{\bm{Q}} \Delta^j_{\bm{Q}}  + \nonumber \\
    && + \frac{1}{2} \sum_{ijkl} 
    \operatorname{Tr} \left[ 
    G_{\bm{k}} \,
    \sigma_0 \otimes \tau_i \,
    G_{\bm{k+Q}} \,
    \sigma_0 \otimes \tau_j \,
    G_{\bm{k}} \,
    \sigma_0 \otimes \tau_k\,
    G_{\bm{k+Q}} \,
    \sigma_0 \otimes \tau_l \,
    \right]
    \bar \Delta^i_{\bm{Q}} \Delta^j_{\bm{Q}} \bar \Delta^k_{\bm{Q}} \Delta^l_{\bm{Q}}, \label{eq:gl_generic_d2h}
\end{eqnarray}
\end{widetext}
where the summation indices run only over $0,z$.
By comparing this expression with the quartic order action (\ref{eq:gl_theory_from_symmetry}), derived from symmetry considerations, we find that
\begin{eqnarray}
    &&a_0 = \operatorname{Tr} \left( 
    G_{\bm{k}}\, 
    \sigma_0\otimes\tau_0\,
    G_{\bm{k+Q}}\, 
    \sigma_0\otimes\tau_0
    \right), \nonumber \\
    &&a_z = \operatorname{Tr} \left( 
    G_{\bm{k}}\,
    \sigma_0 \otimes \tau_z\,
    G_{\bm{k+Q}}\,
    \sigma_0 \otimes \tau_z 
    \right), \nonumber \\ 
    &&b_0 = \frac{1}{2} \operatorname{Tr} \left( 
    G_{\bm{k}}\, 
    \sigma_0\otimes\tau_0\,
    G_{\bm{k+Q}}\, 
    \sigma_0\otimes\tau_0
    \right)^2, \nonumber \\
    &&b_z = \frac{1}{2} \operatorname{Tr} \left( 
    G_{\bm{k}}\, 
    \sigma_0\otimes\tau_z\,
    G_{\bm{k+Q}}\, 
    \sigma_0\otimes\tau_z
    \right)^2.
    \label{eq:quadratic_coeffs_greens_fn}
\end{eqnarray}

\section{\label{sec:in_gap_states}Toy model for in-gap states}
In this section, we present a minimal toy model that captures the essential features of the in-gap states in the presence of the CDW. 
The following $3 \times 3$ Hamiltonian approximately describes the $\Delta^z_{\bm{Q}}$-interaction between the downward-dispersing band of $h_{\bm{k}}$ and the upward-dispersing bands of $h_{\bm{k}+\bm{Q}}$ and $h_{\bm{k}-\bm{Q}}$, along a fixed slice $k_y = a k_x$, for some constant coefficient $a$:
\begin{equation}
    h^{\operatorname{toy}}_{k} = 
    \begin{pmatrix}
        k + q & \Delta & 0 \\
        \Delta & -k & \Delta \\
        0 & \Delta & k - q
    \end{pmatrix},
    \label{eq:toy_ham}
\end{equation}
where $k$ is the momentum parameterizing the slice, $q$ mimics the shift $\bm{Q}$, and $\Delta$ corresponds to the $\Delta^z$-component of the CDW order. Fig. \ref{fig:toy_model} shows the eigenvalues of this Hamiltonian, with the color intensity indicating the spectral weight of the original downward-dispersing band. More precisely, the intensity of each band at a given point $k$ is equal to the squared magnitude of the second component of the corresponding eigenvector at $k$. As $\Delta/q$ increases, the gaps widen and the in-gap states become less pronounced, eventually disappearing completely for $\Delta/q \gg 1$.
\nocite{*}

\bibliography{apssamp}

\end{document}